\documentclass[a4paper,fleqn,usenatbib,useAMS,usedcolumn]{mnras}
\usepackage[T1]{fontenc}
\usepackage{ae,aecompl}
\usepackage{amssymb,amstext,amsfonts}
\usepackage[fleqn]{amsmath}
\usepackage{graphicx}
\usepackage{subfig}
\usepackage{dcolumn}
\usepackage{multirow}
\usepackage{color}
\usepackage{hyperref}

\usepackage{etoolbox}
\makeatletter
\patchcmd\@combinedblfloats{\box\@outputbox}{\unvbox\@outputbox}{}{%
  \errmessage{\noexpand\@combinedblfloats could not be patched}%
}%
\makeatother

\newcommand{\eqnref}[1]{(\ref{eq:#1})}
\newcommand{\figref}[1]{Fig.~\ref{fig:#1}}
\newcommand{\Figref}[1]{Figure~\ref{fig:#1}}
\newcommand{\tabref}[1]{Table~\ref{tab:#1}}
\newcommand{\secref}[1]{Sec.~\ref{sec:#1}}

\DeclareMathOperator{\Dir}{Dir}
\DeclareMathOperator{\DP}{DP}
\DeclareMathOperator{\BetaD}{Beta}
\DeclareMathOperator{\GammaD}{Gamma}
\DeclareMathOperator{\GEM}{GEM}
\DeclareMathOperator{\tr}{tr}

\newcommand{\sub}[1]{\ensuremath{_\mathrm{#1}}}
\newcommand{\super}[1]{\ensuremath{^\mathrm{#1}}}
\newcommand{\dd}{\ensuremath{\mathrm{d}}}

\begin{document}

\title[Volume localisation with gravitational waves]{Dirichlet Process Gaussian-mixture model: An application to localizing coalescing binary neutron stars with gravitational-wave observations}
\author[W.\ Del~Pozzo et al.]
{
\parbox{\textwidth}{W.~Del~Pozzo$^{1,2}$\thanks{E-mail: walter.delpozzo@unipi.it (WDP)},
C.~P.~L.~Berry$^{2}$\thanks{E-mail: cplb@star.sr.bham.ac.uk (CPLB)},
A.~Ghosh$^{3,4}$,
T.~S.~F.~Haines$^{5}$,
L.~P.~Singer$^{6,7}$ 
and A.~Vecchio$^{2}$
}
\vspace{0.2cm}\\
$^{1}$Dipartimento di Fisica ``Enrico Fermi'', Universit\`a di Pisa,  and INFN sezione di Pisa, Pisa I-56127, Italy\\
$^{2}$Institute of Gravitational Wave Astronomy, University of
Birmingham, Edgbaston, Birmingham B15 2TT, United Kingdom\\
$^{3}$Nikhef, Science Park 105, 1098 XG Amsterdam, Netherlands\\
$^{4}$International Centre for Theoretical Sciences, Tata Institute of Fundamental Research, Bangalore 560012, India\\
$^{5}$Department of Computer Science, University of Bath, Claverton Down, Bath, BA2 7AY, United Kingdom\\
$^{6}$Astroparticle Physics Laboratory, NASA Goddard Space Flight Center, 8800 Greenbelt Road, Greenbelt, MD 20771, USA\\
$^{7}$Joint Space-Science Institute, University of Maryland, College Park, MD 20742, USA
}

\date{\today}

\pagerange{\pageref{firstpage}--\pageref{lastpage}} \pubyear{2018}

\maketitle

\label{firstpage}

\begin{abstract}
We reconstruct posterior distributions for the position (sky area and distance) of a simulated set of binary neutron-star gravitational-waves signals observed with Advanced LIGO and Advanced Virgo. 
We use a Dirichlet Process Gaussian-mixture model, a fully Bayesian non-parametric method that can be used to estimate probability density functions with a flexible set of assumptions.
The ability to reliably reconstruct the source position is important for multimessenger astronomy, as recently demonstrated with GW170817. We show that for detector networks comparable to the early operation of Advanced LIGO and Advanced Virgo, typical localization volumes are $\sim10^4$--$10^5~\mathrm{Mpc^3}$ corresponding to $\sim10^2$--$10^3$ potential host galaxies. The localization volume is a strong function of the network signal-to-noise ratio, scaling roughly $\propto \varrho\sub{net}^{-6}$. Fractional localizations improve with the addition of further detectors to the network. Our Dirichlet Process Gaussian-mixture model can be adopted for localizing events detected during future gravitational-wave observing runs, and used to facilitate prompt multimessenger follow-up.
\end{abstract}


\section{Introduction}

Bayesian inference is frequently used in astronomy as a means of combining new data with prior knowledge to construct a better model for our understanding of astronomical systems. Our state of knowledge about the values of a system's parameters is encoded in a probability distribution. An efficient and effective means of mapping a probability distribution is by using a stochastic sampling algorithm, such as nested sampling \citep{Skilling2006} or Markov-chain Monte Carlo \citep[chapter 12]{Gregory2005}. These explore parameter space and, in so doing, return a set of samples randomly drawn from the desired probability distribution. These samples can be used to calculate summary statistics such as expectation values; however, for some applications it is desirable to have a smooth probability density function. This leaves the question of converting a discrete set of samples into a continuous probability density function.

The crudest means of reconstructing a probability density function is by creating a set of bins and counting the number of samples that fall in each. This is extremely difficult to do robustly: bins must be sufficiently small to resolve the features of the distribution (and avoid introducing artefacts from the quantization) but still large enough that they contain sufficient samples to provide a fair estimate of the underlying probability density at that location. It is almost impossible to do this using a single bin size; in practice we must adapt to the shape of the distribution, which is not usually known beforehand.

In this paper we explain an algorithm, using Dirichlet processes (DPs) to build a Gaussian-mixture model (DPGMM), that can be used to build probability distributions from a set of samples. We specialise to the question of inferring the (three-dimensional) location of an astronomical system; however, the algorithm may be generalised for working with different parameter spaces. Our DPGMM can be used to efficiently combine the three-dimensional probability distribution with a galaxy catalogue to produce a list of most probable host galaxies.

This work originates from the field of gravitational-wave astronomy. The new generation of detectors began operation in September 2015 \citep{Abbott2017e}, with the first observing run (O1) of Advanced Laser Interferometer Gravitational-wave Observatory \citep[aLIGO;][]{Aasi2015}. This run yielded the first observations of binary black hole coalescences, GW150914 \citep{Abbott2016}, GW151226 \citep{Abbott2016e} and potentially LVT151012 \citep{Abbott2016g,Abbott2016d}. The second observing run (O2) began in November 2016, with Advanced Virgo \citep[AdV;][]{Acernese2015} joining for the final month of August 2017. The extension of the gravitational-wave detector network to include additional observatories improves the prospects for localizing the source on the sky \citep[e.g.,][]{Abbott2017e,Singer2014,Gaebel2017}. O2 saw further binary black hole detections, GW170104 \citep{Abbott2017}, GW170608 \citep{Abbott2017f} and GW170814 \citep{Abbott2017d}, as well as the first binary neutron star (BNS) detection, GW170817 \citep{Abbott2017c}. The complete results of O2 are yet to be announced.

Gravitational-wave observations do not pinpoint the source of transient signals, instead the source location is inferred probabilistically. 
The source location is of paramount importance for identifying a multimessenger counterpart: both for targeting follow-up observations and for establishing that a candidate counterpart is associated with the gravitational-wave source.\footnote{It may be sufficient to associate a gravitational-wave signal with a gamma-ray burst by time coincidence alone, since both are short lived \citep[cf.][]{Aasi2014f,Aasi2014e,Abbott2016o}, but additional spatial coincidence gives greater confidence \citep[cf.][]{Abbott2017g}.} Extensive electromagnetic and neutrino follow-up has been performed for the binary black hole detections \citep[e.g.,][]{Abbott2016a,Adrian-Martinez2016,Albert2017}, with no conclusive counterpart yet found. This is not surprising. BNSs are the more promising source for counterparts \citep[e.g.,][]{Metzger2012,Piran2013}, and GW170817 was accompanied by detections across the electromagnetic spectrum \citep{Abbott2017b}. A short gamma-ray burst, GRB\,170817A, was observed independently of the gravitational-wave localization \citep{Goldstein2017,Savchenko2017}, but the (three-dimensional) localization from gravitational-wave observations was crucial for identification of a kilonova counterpart \citep{Coulter2017,Soares-Santos2017,Valenti2017,Arcavi2017,Tanvir2017,Lipunov2017}. Multimessenger observations give a range of insights, such as testing the speed of gravity \citep{Abbott2017g}; exploring the host environment and formation history of merging compact binaries \citep{Abbott2017j,Blanchard2017,Im2017,Levan2017,Pan2017}, and estimation of the Hubble constant \citep{Abbott2017h,Guidorzi2017}. 
The question of sky-localization potential for a realistic astrophysical population of BNS systems has been investigated in \citet{Singer2014} and \citet{Berry2014}. For the early observing runs, localizations were typically of the order of hundreds of square degrees, making follow-up observations challenging. The probability of observing a counterpart can be enhanced by using galaxy catalogues to pick out the most likely locations \citep{Hanna2014,Fan2014}; including information on the distance of the source can significantly aid this process \citep{Nissanke2012,Gehrels2015,Singer2016}. 

Even without observing a counterpart, inferring the (three-dimensional) location of gravitational-wave sources is useful. Comparing posterior distributions on location with galaxy catalogues makes it possible to assign a probability that a signal originated from a particular galaxy. Comparing the luminosity distance from the gravitational-wave observation with the redshift measurements for the galaxies gives a measure of the Hubble constant \citep{Schutz1986}. Combining results from a few tens of observations from the advanced-detector network could measure the Hubble constant to an accuracy of $\sim5\%$ at $95\%$ credibility \citep{DelPozzo2012,Chen2017}. This is comparable to existing constraints from the \textit{Hubble Space Telescope} Key Project \citep{Freedman2001}, and inferior to current results from the \textit{Planck} cosmic microwave background observations \citep{Ade2015}, the SH0ES type Ia supernovae survey \citep{Riess2016,Riess2018}, or from the weak lensing measurements (combined with baryonic acoustic oscillation and Big Bang nucleosynthesis data) from the Dark Energy Survey \citep{Abbott2017i}. However, the gravitational-wave measurement is independent of the usual systematics, making it a valuable check.

While the primary purpose of this work is to document our implementation of a DPGMM for gravitational-wave source localization, and to demonstrate its effectiveness, the techniques described are of general applicability, and could be of interest for a wide range of problems. We begin in \secref{Dirichlet} with background material on DPs and the DPGMMs; those only interested in our results may skip this section. We apply the DPGMM to reconstruct the position posterior probabilities densities of a set of simulated BNS signals. We use the (well studied) catalogue of results generated to model the expected early operation of the advanced-detector network presented in \citet{Singer2014} and \citet{Berry2014}; this is described in \secref{data}. In \secref{results}, we present our results for the source localization. Our reconstructed three-dimensional posteriors indicate that BNSs could be localised to $\sim10^4$--$10^5~\mathrm{Mpc}$ during the early runs of the advanced-detector era, assuming perfect detector calibration \citep[cf.][]{Singer2016}. The introduction of more detectors will improve both two-dimensional and three-dimensional localization, and so the probability of successfully identifying multimessenger counterparts to the gravitational-wave signal.

\section{Use of Dirichlet processes}\label{sec:Dirichlet}

\subsection{Posterior distributions}

In many fields of astronomy and astrophysics, one of the main challenges is to be able to accurately measure the physical parameters of interest and consequently make reliable statements about the systems that have been observed. Given a set of observations and a model, one must infer the values of the parameters. The dimensionality of parameter space is frequently large, necessitating the use of stochastic samplers for exploration \citep[chapter 29]{MacKay2003}. For making reliable inferences about compact binary coalescences (the inspiral and merger of neutron star--neutron star, neutron star--black hole and black hole--black hole binaries), the LIGO Scientific and Virgo Collaborations (LVC) have devoted significant time and effort to develop \textsc{LALInference} \citep{Veitch2014}, a suite of programs that are part of the LVC Algorithm Library (LAL).\footnote{In addition to the stochastic sampling algorithms of \textsc{LALInference}, localization of BNSs can also be performed using \textsc{bayestar} \citep{Singer2015a}, a more expedient algorithm, which we do not consider here.} Other fields have equivalent specialised codes, such as \textsc{CosmoMC} for cosmic microwave background (and other cosmological observations) analysis \citep{Lewis2002} or \textsc{TempoNest} for pulsar timing \citep{Lentati2013}, or may use general samplers like \textsc{emcee} \citep{Foreman-Mackey2013}. The output of any of these is a list of independent samples drawn from the posterior probability distribution of all relevant parameters. These samples can then be used to reconstruct information about the parameters of interest.

For some applications it is desirable to have a smooth estimate of the posterior probability density functions. For example, in our case, we will use the probability density functions to (i) calculate credible volumes to check and summarise our reconstructed localizations, and (ii) correlate with galaxy catalogues to find the most probable host galaxies. The discrete nature of the samples can make computing the probability density function difficult. To address this problem, various techniques have been developed; the most common ones are histogramming and kernel density estimation (KDE). Both techniques can be effective when the shape of the posterior distribution function is simple or when the number of samples is large; however, when the number of samples is small, different choices of the bin size for histograms or of the kernel width for KDE can yield distorted results that depend on the actual choice of these parameters. Aware of these limitations, an alternative technique based on constructing a $k$-dimensional tree has been suggested for the estimation of credible regions in the two-dimensional sky plane \citep{Sidery2014}.\footnote{This uses a two-step algorithm to ensure unbiased results \citep{Berry2013a,Sidery2014a}.} This method successfully estimates the sky position, but since it must tile the region of interest with rectangular leaves, its applicability is still limited to simple distributions or large sample numbers. In this paper, we present a Bayesian non-parametric technique based on the DP, that can be used on any set of posterior samples.

Our method is routinely used in different fields, e.g., in the context of unsupervised pattern recognition and non-parametric density estimation, but, to the best of the authors' knowledge, it is largely unknown to the astrophysical and gravitational-wave communities. A thorough introduction can be found in the compendium \citet{Hjort2010}; we give a short overview in this section. We begin by introducing the finite-dimensional version of the DP, which is the Dirichlet distribution (\secref{DD}). We then describe the DP itself (\secref{DP}) and how it can be used to reconstruct a probability density function using a Gaussian-mixture model (\secref{DPGMM}). Some specifics of our implementation of the DPGMM are described in \secref{implement}. 

\subsection{The Dirichlet distribution}\label{sec:DD}

Consider a random experiment which can give a finite number of outcomes, and imagine that we are only interested in registering the class of the outcome. For example, we may be interested in a coin toss where the outcome is either heads or tails, classifying a gravitational-wave source as a BNS, a neutron star--black hole or a binary black hole system, or registering the number of samples that fall inside a bin in order to construct a histogram. If we have $k$ categories, after $N$ samples, the likelihood of the observations is given by the multinomial distribution
\begin{equation}
p(n_1,\ldots,n_k|q_1,\ldots,q_k) = \frac{N!}{n_1!\ldots n_k!} \prod_{i=1}^k q_i^{n_i}\,,
\label{eq:multinomial}
\end{equation}
where $n_i$ is the number of samples in the $i$-th category, so $N \equiv \sum_{i=1}^k n_i$, and $q_i$ is the corresponding probability for a sample to be in that category. In a frequentist context, these probabilities can be estimated from the observed frequencies of each outcome, which becomes exact as $N$ tends to infinity. However, there is nothing stopping us from applying Bayes theorem and asking: ``given the observed samples, how plausible are the inferred probabilities?'' \citep[chapter 18]{Jaynes2003}. In other words, given the observed data, one can assign a probability distribution to the probabilities for each category.

To infer the probabilities $\boldsymbol{q} \equiv \{q_i\}$ given the observed counts $\boldsymbol{n} \equiv \{n_i\}$ we can make use of Bayes' theorem, 
\begin{equation}
p(\boldsymbol{q}|\boldsymbol{n}) = \frac{p(\boldsymbol{n}|\boldsymbol{q})p(\boldsymbol{q})}{\int \dd\boldsymbol{q}\,p(\boldsymbol{n}|\boldsymbol{q})p(\boldsymbol{q})}\,,
\end{equation}
where $p(\boldsymbol{n}|\boldsymbol{q})$ is the likelihood defined in \eqnref{multinomial} and $p(\boldsymbol{q})$ is the prior distribution on the probabilities $\boldsymbol{q}$. To complete the inference, we only need to select an appropriate prior.

When we are interested in estimating the probability mass function from the observation of a discrete set of samples, a prior is required for the problem to be well
posed. Without assigning a prior, estimating a probability density from a histogram can be, in some cases, troublesome. For instance, if one of the bins has been assigned no samples, the probability assigned to that particular bin will always be zero. Inclusion of a suitable prior circumvents this issue, since it allows for a non-zero probability in each bin even without any observations (the role of the prior is to say that we expect that it is possible for a sample to be in each category). Therefore, we obtain sensible results from our inference, even when we have few samples.

A common choice for a prior in this situation is the Dirichlet distribution. As we shall see, the Dirichlet distribution has several convenient properties that allow it to be tailored to match our prior expectations. One advantage of using the Dirichlet distribution is that it is \emph{conjugate} to the multinomial distribution \citep[chapter 3]{Raiffa1961}. This means that if we use a Dirichlet distribution as a prior with our multinomial likelihood, our posterior will also be a Dirichlet distribution (which can then be used as the prior for our next set of observations). This invariance under the inclusion of new data means that our inferences form a never-ending chain of Dirichlet distributions, which greatly simplifies computation and interpretation of results \citep[section 2.4]{Gelman2014}.

The Dirichlet distribution is defined as
\begin{equation}
\Dir(\boldsymbol{q}|\boldsymbol{a{}}) = \frac{\Gamma (A)}{\prod_{i=1}^k \Gamma(a{}_i)} \prod_{i=1}^k q_i^{a{}_i-1} \qquad
\qquad \{a{}_i > 0\}\,,
\end{equation}
where $\Gamma$ is the gamma function, $\boldsymbol{a{}} \equiv \{a{}_1,\ldots,a{}_k\}$ are the concentration parameters, which control the shape of the distribution; $A \equiv \sum_{i=1}^{k} a{}_i$, and the probabilities $\boldsymbol{q}$ are normalised such that
\begin{equation}
\sum_{i=i}^k q_i =1 \,.
\end{equation} 
With a Dirichlet prior, the posterior distribution for the probabilities $\boldsymbol{q}$ given some data counts $\boldsymbol{n}$ is then
\begin{equation}
p(\boldsymbol{q}|\boldsymbol{n}) = \Dir(\boldsymbol{q}|\boldsymbol{a{}}+\boldsymbol{n}) \,.
\end{equation}
Hence, we can consider $\boldsymbol{a{}}$ as the set of prior counts for each category observed before our current observation set; since these are non-zero, we ensure that even when we have no samples in a bin, its probability is not zero. In general, for $\boldsymbol{q} \sim \Dir(\boldsymbol{a{}}+\boldsymbol{n})$, the expectation (mean) value of probability $q_i$ is
\begin{equation}
\label{eq:qbar-general}
\bar{q}_i = \frac{a{}_i + n_i}{A + N}\,;
\end{equation}
thus, in the limit of $n_i \gg a{}_i$, such that the likelihood dominates over the prior, we recover the intuitive frequentist result $n_i/N$.

The Dirichlet distribution is a practical density estimator for discrete probability distributions. When we are beginning our inferences, we are typically starting from a state of ignorance: we do not prefer any one category over another and therefore must assign each equal probability. The corresponding uninformative choice of the Dirichlet distribution has \citep[section 3.4]{Gelman2014}\footnote{Setting the $a{}_i$ to any constant will result in a uniform distribution. The choice of $a{}_i = 1$ has the attractive property of corresponding to a prior weight of each bin having a single count. Using a larger value gives a stronger prior on the distribution being uniform, and more samples need to be collected before the inferred distribution will significantly deviate from this.}
\begin{equation}
a{}_i = 1\,.
\end{equation}
Following collection of the samples, application of Bayes' theorem with this prior gives an expectation value
\begin{equation}
\bar{q}_i = \frac{n_i + 1}{N + k}\,.
\end{equation}
For the case of two possible outcomes, this yields Laplace's rule of succession (\citealt[chapter 18]{Jaynes2003}; \citealt[section 3.2]{MacKay2003}). The modal value (maximum a posteriori estimate) for probability $q_i$ is
\begin{equation}
\hat{q}_i = \frac{n_i}{N}\,,
\end{equation}
agreeing with the frequentist result.

Having established the properties of the Dirichlet distribution, we now consider its infinite-dimensional generalization, the DP.

\subsection{The Dirichlet process}\label{sec:DP}

The Dirichlet process (DP) was introduced in \citet{Ferguson1973}. It is a stochastic process that generalises the Dirichlet distribution to infinite dimensions and can be used to set a prior on unknown distributions. While a draw from the Dirichlet distribution is a discrete distribution of finite length, a draw from the DP is a discrete distribution of infinite length. It is a probability distribution for other probability distributions; this additional freedom allows us to dispense with the need to specify bins. For a historical introduction to the DP and its properties, see \citet{Gupta2001}. 

To define a DP, let us consider a probability distribution $G$ over the parameter space $\Theta$.\footnote{For our application, $\Theta$ can be interpreted as the space of means and covariances that define our smoothing kernels (see \secref{DPGMM}).} We use $\vartheta$ to denote an element or collection of elements of $\Theta$, with $G(\vartheta)$ the corresponding probability (density). For $G$ to be DP distributed we require that for \emph{any} set of partitions $\vartheta_1,\ldots,\vartheta_k$ of $\Theta$ (these could represent histogram bins), the vector $\boldsymbol{G} = (G(\vartheta_1),\ldots,G(\vartheta_k))$ is distributed according to a Dirichlet distribution. Introducing a base distribution $H$ over $\Theta$ with $\boldsymbol{H} = (H(\vartheta_1),\ldots,H(\vartheta_k))$, and a (positive, real) concentration parameter $a{}$, we have that
\begin{equation}
\boldsymbol{G} \sim \Dir(a{} \boldsymbol{H})\,,
\label{eq:DP-DD}
\end{equation}
and we say that $G$ is DP distributed with base distribution (or base measure) $H$ and concentration parameter $a{}$,
\begin{equation}
G \sim \DP(a{}, H)\,.
\end{equation}
Intuitively, $H$ can be thought as the mean of the DP: distributions are drawn from around $H$ such that the expectation value is $\bar{G}(\vartheta) = H(\vartheta)$. The concentration parameter $a{}$ plays the role of the inverse variance of the DP, controlling how the samples are distributed across $\Theta$: in the limit of $a{} \rightarrow 0$, the draws are all clustered at a single, random $\vartheta$, while in the limit of $a{} \rightarrow \infty$ the draws follow exactly the base distribution \citep[section 23.2]{Gelman2014}.\footnote{In \eqnref{DP-DD}, the Dirichlet distribution only depends upon the product $a{} \boldsymbol{H}$, but the potential degeneracy between the magnitude of $a{}$ and $\boldsymbol{H}$ is broken by requiring that $H$ is normalised to unity.} When a DP is used for inference, the concentration parameter controls the strength of the prior, with a larger value keeping us closer to our initial expectation of a distribution like $H$, in a similar way to how $\boldsymbol{a{}}$ sets the prior strength in a Dirichlet distribution \citep[cf.][section 3.3.4]{Raiffa1961}.

The DP has a similar conjugacy property to the Dirichlet distribution. Let us imagine that we have collected $N$ observations $\zeta_i \sim G$, where $i$ runs from $1$ to $N$. If our prior is $G \sim \DP(a{}, H)$, then our posterior would be \citep[section 23.2]{Gelman2014}
\begin{equation}
G \sim \DP\left(a{}+N, \frac{a{}}{a{}+N}H(\vartheta) + \frac{1}{a{}+N}\sum_{i=1}^N \delta(\vartheta - \zeta_i)\right)\,.
\end{equation}
From this, we can obtain the posterior expectation of $G$, which is now our best prediction for future observations \citep{Blei2006,Teh2010},
\begin{equation}
\bar{G}(\vartheta) = \frac{a{}}{a{} + N}H(\vartheta) + \frac{1}{a{} + N}\sum_{i=1}^{N}\delta(\vartheta - \zeta_i)\,.
\end{equation}
The form is analogous to that in \eqnref{qbar-general}. We now need to know how to make use of the posterior DP.

Samples from a DP are a weighted sum of point probability masses, and they can be constructed in several ways (such as the Blackwell--MacQueen urn scheme, Chinese restaurant process or stick-breaking construction), each emphasising a different property of the DP \citep{Teh2010}. We use the stick-breaking construction, where a sample from a DP $G \sim \DP(a{},H)$ can be represented as \citep{Sethuraman1994}
\begin{equation}
\label{eq:dp-sample}
G(\vartheta) = \sum_{i=1}^\infty w_i \delta(\vartheta - \zeta_i)\,,
\end{equation}
where
\begin{align}
\label{eq:GEMw}
w_j  = {} & \beta_j\prod_{i=1}^{j-1}(1-\beta_i)\,,\\
\label{eq:GEMbeta}
\beta_j \sim {} & \BetaD(1,a{})\,,\\
\zeta_i \sim {} & H\,.
\end{align}
Here, the beta distribution is
\begin{equation}
\BetaD(\beta|a,b) = \frac{\Gamma(a+b)}{\Gamma(a)\Gamma(b)} \beta^{a-1}(1-\beta)^{b-1}\,; 
\end{equation}
it is the binomial specialisation of the Dirichlet distribution. For brevity, we can combine \eqnref{GEMw} and \eqnref{GEMbeta}, and denote $\boldsymbol{w} \equiv \{w_j\}$ as being constructed following the Griffiths--Engen--McCloskey (GEM) distribution \citep[chapter 3]{Pitman2006},
\begin{equation}
\boldsymbol{w} \sim \GEM(a{})\,.
\end{equation}
Since a sample $G(\vartheta)$ from a DP can be interpreted as a collection of point probability masses, it is a discrete distribution; $G(\vartheta)$ has no density, but is instead atomic. Consequently, samples from a DP cannot be used directly to describe continuous distributions. Nevertheless, DPs are commonly used for non-parametric density estimation by using draws from a DP to define a set of kernel functions \citep{Lo1984,Escobar1995}. We use a Gaussian-mixture model to reconstruct our inferred probability distribution as described in the next section.

\subsection{The Gaussian-mixture model}\label{sec:DPGMM}

To build a continuous probability density function from our DP draws, we use a mixture of smoothing kernel functions. Let us introduce $\mathcal{K}(\xi|\vartheta)$ as the family of kernel functions indexed by $\vartheta$. Using our DP-distributed $G$, we can build a non-parametric probability density for $\xi$ according to \citep[section 23.3]{Gelman2014}
\begin{equation}
p(\xi) = \int \dd\vartheta\, \mathcal{K}(\xi|\vartheta)G(\vartheta)\,.
\end{equation}
This can be turned into a sum, an infinite mixture of kernels, using \eqnref{dp-sample}.

The common choice for the kernel function is a multivariate Gaussian
\begin{equation}
\mathcal{K}(\xi|\vartheta) \equiv \mathcal{N}(\xi| \mu,S^{-1})\,,
\end{equation}
where $\mu$ is the (multidimensional) mean and $S$ is the precision matrix (the inverse of the covariance matrix). This choice defines the Dirichlet Process Gaussian-mixture model (DPGMM); we describe the distribution for $\xi$ as being made up of an infinite mixture of Gaussian clusters, each with their own mean and covariance. The mean and precision matrix are learned from the data when fitting the DP model.

To define the DP for $\mu$ and $S$, we must specify a base distribution. It is common practice to use conjugate priors for these applications, to exploit their useful properties. Different choices are possible \citep{Gorur2010}, but at the price of losing the conjugacy property and therefore complicating the analysis substantially. The conjugate prior of a multivariate Gaussian distribution with unknown mean and precision matrix is the normal--Wishart distribution \citep[cf.][]{Escobar1995}
\begin{equation}\label{eq:nw}
\mathcal{NW}(\mu,S|\mu_0,\rho,\Lambda,\nu) = \mathcal{N}\left(\mu\middle|\mu_0,(\rho S)^{-1}\right) \mathcal{W}(S|\Lambda,\nu)\,.
\end{equation}
Here, the Wishart distribution with $\nu$ degrees of freedom is
\begin{align}
\mathcal{W}(S|\Lambda,\nu) = {} & \frac{|\Lambda|^{-\nu/2}}{2^{\nu m/2}\pi^{m(m-1)/4}}\left[\prod_{i=1}^{m}\Gamma\left(\frac{\nu+1-i}{2}\right)\right]^{-1} \nonumber\\*
 & {} \times |S|^{-(\nu+m+1)/2}\exp\left[-\frac{1}{2}\tr(\Lambda^{-1} S)\right]\,,
\end{align}
where $S$ and $\Lambda$ are positive-definite $m \times m$ matrices, and the expectation value is $\bar{S} = \nu \Lambda$. The normal--Wishart distribution introduces hyperparameters $\mu_0$ (the expected value of the mean), $\rho$ (a scale factor), $\Lambda$ (a prior for the precision matrix) and $\nu$ (the number of degrees of freedom); these are common to all mixture components, expressing the belief that component parameters should be members of a single family. We choose the parameters of the normal--Wishart distribution to be the mean and precision of the observed samples, the scale factor to be equal to the requested resolution (see \secref{implement} for further details), and the number of degrees of freedom to be equal to the dimensionality of the problem plus two (this ensures that the distribution is well conditioned).

Due to its conjugacy to the multivariate Gaussian, choosing $\mathcal{NW}(\mu,S|\mu_0,\rho,\Lambda,\nu)$ as the base distribution for the DP, it is possible to marginalize out analytically the multivariate Gaussian parameters and obtain the non-parametric density estimate as a mixture of multivariate Student-$t$ distributions.\footnote{The normal distribution is a limiting case of the Student-$t$ distribution.} 

In addition to the base distribution, we also need a concentration parameter for our DP. This too can be updated from the data, but we must specify a prior distribution for it. We use a gamma distribution \citep{Escobar1995}, specifically $a{} \sim \mathrm{Gamma}(1,1)$. The gamma distribution is given by
\begin{equation}
\GammaD(a|b,c) = \frac{c^b}{\Gamma(b)} x^{b-1} \exp(-cx)\,;
\end{equation}
it is the univariate specialization of the Wishart distribution. It is especially convenient as it is conjugate to the beta distribution used in \eqnref{GEMbeta} \citep{Blei2006}. The prior expectation is $\bar{a} = 1$ \citep[cf.][section 23.3]{Gelman2014}. 

Combining everything together, the prior DPGMM is assembled as
\begin{align}
a{} \sim {} & \GammaD(1,1)\,, \\
\boldsymbol{w} \sim {} & \mathrm{GEM}(a{})\,, \\
\mu_i,\,S_i \sim {} & \mathcal{NW}\left(\mu,S\middle|\mu_0,\rho,\Lambda,\nu\right)\,, \\
\xi \sim {} & \sum_{i=1}^\infty w_i \mathcal{N}(\mu_i,S_i^{-1})\,.
\end{align}
We first calculate hyperparameters (concentration and base distirbution) to specify our DP; this determines parameters that describe a mixture of Gaussian kernels, and the sum of this mixture gives the distribution of the observed parameters $\xi$ (in \secref{implement} we describe how $\xi$ is a set of three-dimensional position coordinates). Given a set of data (particular realizations of $\xi$), we now have to solve the inverse problem to find its posterior probability density.

DPGMMs can be explored using Gibbs sampling \citep{Neal2000,Rasmussen2000}; however, we use the variational algorithm introduced in \citet{Blei2006} with the capping method described in \citet{Kurihara2007}. We make use of the publicly available implementation developed by one of the authors \citep[previous applications include background subtraction;][]{Haines2012,Haines2014}.\footnote{The \textsc{dpgmm} module is available from \href{https://github.com/thaines/helit/}{github.com/thaines/helit/}.} Our choice of implementation allows the number of components in the DPGMM to grow without limit until the best fitting model is found; this finite number of components is then used as our estimate for the posertior probability density. The multivariate normal mean vector and covariance matrix are set by maximising the likelihood of the observed data vector $\xi$, given the number of components to which data have been assigned, see equation~(17) in \citet{Gorur2010}.

\subsection{Implementation for gravitational-wave data}\label{sec:implement}

We are interested in reconstructing posterior probability densities from a set of samples as calculated by a stochastic sampling algorithm \citep{Veitch2014}. To do so, we have adopted the algorithm presented in the previous subsection, specialised to the problem of estimating the posterior probability density for the distance $D$, right ascension $\alpha$ and declination $\delta$.\footnote{We neglect the effects of cosmology and so do not distinguish between different distances; the furthest source we consider is at a (luminosity) distance of $269~\mathrm{Mpc}$, which corresponds to a redshift of less than $0.07$ assuming standard cosmology \citep{Ade2015}.}

Since the DPGMM is not designed to deal with periodic coordinates, we perform our analysis in Cartesian coordinates; we transform $\{D,\alpha,\delta\}$ into $\{x,y,z\}$ following the transformation 
\begin{align}
x = {} & D \cos(\alpha)\cos(\delta)\,, \\
y = {} & D \sin(\alpha)\cos(\delta)\,, \\
z = {} & D \sin(\delta)\,.
\end{align} 
It is these Cartesian-space samples that define our observations $\xi$, and we use their mean and inverse covariance to specify the hyperparameters of the normal--Wishart distribution \eqnref{nw}. We apply the variational method of \citet{Blei2006}, as described in \secref{DPGMM}, to obtain the probability density $p(x,y,z|\boldsymbol{w},\mu,S^{-1})$. We transform back into $\{D,\alpha,\delta\}$-space using the Jacobian of the coordinate transformation,
\begin{equation}
p(D,\alpha,\delta|\boldsymbol{w},\mu,S^{-1}) = p(x,y,z|\boldsymbol{w},\mu,S^{-1})\left\|\frac{\partial(x,y,z)}{\partial(D,\alpha,\delta)}\right\|\,,
\end{equation}
where
\begin{equation}
\left\|\frac{\partial(x,y,z)}{\partial(D,\alpha,\delta)}\right\| = D^2\cos(\delta)\,.
\end{equation}
We then obtain the non-parametric posterior density estimate by marginalising away $\mu$ and $S$ analytically, thanks to the choice of conjugate priors. 

Once we have obtained $p(D,\alpha,\delta)$, we can use it for making statements about the probable location. For example, we can compute credible volumes by evaluating the model over a three-dimensional grid spanning the whole volume under consideration. By default, we use a uniform $\{D,\alpha,\delta\}$ grid which is $50 \times 1440 \times 720$. This is by far the most computationally expensive step in our analysis, taking on the order of $\sim1~\mathrm{hr}$.\footnote{Across all data sets, the median run time is $2900~\mathrm{s}$ and the central $90\%$ range is $20$--$4340~\mathrm{s}$ using eight CPU cores.} 
Possibilities for optimising this, such as using an adaptive grid, will be investigated in the future. Once the density function has been evaluated over the grid, we sort each of the grid points according to their probability, compute the cumulative distribution and then find the set of points having a probability equal to the requested credible level. Two-dimensional posterior distributions for sky position, as well as one-dimensional posterior distributions for distance, are then obtained by numerical marginalisation of the original three-dimensional distribution. Credible regions and intervals in the lower-dimensional spaces are obtained in the same way as their three-dimensional counterparts. 
As we explain in \secref{EM}, we can also use $p(D,\alpha,\delta)$ directly, without computing credible volumes, together with galaxy catalogues to produce a list of most probable source galaxies.

\section{Simulation}\label{sec:data}

To demonstrate the effectiveness of the DPGMM at estimating probability density functions, we consider the problem of reconstructing the posterior distribution for the position of a (simulated) BNS coalescence, as would be observed in the early advanced gravitational-wave detector era (similar to during O1 and O2). The (three-dimensional) position is an illustrative test case since it gives an indication of how the technique functions in multiple dimensions, while still being easy to visualise. However, our main motivation for considering the position is the desire to be able to reliably reconstruct the location of a gravitational-wave source following a detection for the purposes of electromagnetic or neutrino follow-up \citep[e.g.,][]{Abbott2017e,Abbott2016a,Adrian-Martinez2016,Albert2017,Abbott2017b,Albert2017a}.

We make use of the data presented in \citet{Singer2014} and \citet{Berry2014}. These consider two observing scenarios in anticipation of the early operation of the advanced detector network. The first scenario considers the two-detector network of LIGO Hanford and LIGO Livingston, with sensitivities similar to what was expected for O1; the second considers the three-detector network including AdV, with sensitivities similar to what was expected in O2; we refer to these scenarios as HL and HLV respectively.\footnote{The HL and HLV scenarios are the 2015 and 2016 scenarios of \citet{Singer2014}, respectively.} \citet{Singer2014} simulated two months of observations from each scenario, while \citet{Berry2014} only considered the HL scenario, but used more realistic noise, including non-Gaussianity as seen in the sixth science (S6) run of initial LIGO \citep{Aasi2014c}. 
The detectors' duty cycles are assumed to be $80\%$ \citep[cf.][]{Abbott2017e}, such that in the HLV scenario there are three-detector observations for $51.2\%$ of the time and two-detector observations for $38.4\%$ of the time. The assumed HL sensitivity was slightly less than actually achieved in O1, the assumed BNS detection range was $\sim55~\mathrm{Mpc}$ compared with the achieved range of $\sim70~\mathrm{Mpc}$ \citep{Abbott2016m}; conversely, the assumed HLV sensitivity was better than achieved for the majority of O2 \citep{Abbott2017,Abbott2017d}. However, these data sets provide a qualitative illustration of what can be achieved during the early observing runs of the aLIGO--AdV network.

We refer to the \citet{Singer2014} results as HL Gaussian and HLV Gaussian, since the detector noise is Gaussian, and the \citet{Berry2014} results as HL recoloured, because the noise is recoloured S6 noise.\footnote{The recolouring process consists of first whitening the noise (removing the colour), removing initial LIGO's frequency dependence, and then passing the noise through a filter (reintroducing colour) so that, on average, it has the aLIGO spectral density. This ensures the noise contains realistic non-stationary and non-Gaussian features, although these are not identical to those in the advanced detectors.} Both share the same catalogue of sources, an astrophysically motivated population of BNSs. The neutron-star masses were chosen to be uniformly distributed between $1.2 M_{\odot}$ and $1.6 M_{\odot}$; the sources were distributed uniformly in co-moving volume and on the polarisation--inclination two-sphere, and each neutron star was given a randomly oriented spin with a uniformly distributed magnitude up to a maximum $\chi\sub{max} = 0.05$;\footnote{The dimensionless spin magnitude is $\chi\sub{max} = c|\boldsymbol{S}|/Gm^2$, where $|\boldsymbol{S}|$ is the modulus of the star's spin angular momentum vector and $m$ is its mass. The limit $\chi\sub{max} = 0.05$ matches that assumed for the low-spin prior used in the analysis of GW170817 \citep{Abbott2017c}.} these ranges cover the observed population of BNSs \citep[e.g.,][]{Mandel2009,Ozel2012,Kiziltan2013,Abbott2017c}. Further details about the simulation can be found in \citet{Singer2014}.

The simulated data were treated as real signals would be, first being passed through the \textsc{GstLAL} detection pipeline \citep{Cannon2012}. On account of the difference in noise, slightly different detection criteria were used in \citet{Singer2014} and \citet{Berry2014}, the former using a cut in the network signal-to-noise ratio (SNR) of $\varrho\sub{net} = 12$ and the latter using a false-alarm rate (FAR) threshold of $10^{-2}~\mathrm{yr^{-1}}$. Although broadly consistent, this difference results in the inclusion of additional low SNR ($\varrho\sub{net} \approx 10$--$12$) events for the FAR-only cut.

Parameter-estimation codes are run on detections \citep{Abbott2016b,Abbott2016d,Abbott2017c}, and we use the posterior samples generated by \textsc{LALInference} \citep{Veitch2014}. This analysis, for expediency, did not include the effects of the spins of the neutron stars; this does not influence our results, as spins do not impact the inferred localization when they are small as for our BNSs \citep{Farr2015}. The results also do not include the effects of uncertainty in the detector calibration. Initial results from aLIGO had $10\%$ uncertainty in amplitude and $10~\mathrm{deg}$ uncertainty in phase \citep{Abbott2016c}, and this increased uncertainty in sky localization by a factor of $\sim3$--$4$ for GW150914 \citep{Abbott2016b}; however, the accuracy of calibration had been improved by the end of the run, such that its effects only increased the uncertainty in GW150914's sky localization by a factor of $\sim1.3$--$1.5$, and made negligible difference for the localization of LVT151012, GW151226 or GW170104 \citep{Abbott2016d,Abbott2017}.

Sky-localization accuracy and the distance estimation have been considered previously, and the three-dimensional localization remains an active area of research. Prospects for improving electromagnetic follow-up using a low-latency three-dimensional localization are discussed in \citet{Singer2016}. The approach outlined in \citet{Singer2016} was used during O2 to provide prompt localizations using the \textsc{bayestar} algorithm \citep{Singer2015a}. It approximates the posterior distribution along a line of sight using an ansatz distribution, which assumes that the likelihood is Gaussian \citep[cf.][]{Cutler1994}. The resulting probability distributions can be efficiently communicated as a list of moments for pixels describing different lines of sight. At higher latencies, three-dimensional localizations were provided in O2 using the posterior samples from \textsc{LALInference}. These were post-processed using a clustering KDE algorithm, which is an updated version of the code used to construct the two-dimensional localizations in \citet{Singer2014} and \citet{Berry2014}.\footnote{The KDE clustering algorithm, and accompanying documentation, is available from \href{http://github.com/farr/skyarea}{github.com/farr/skyarea}.} This code performs the KDE in Cartesian coordinates. The resulting distribution is then simplified, so that the results can be communicated using the same summary statistics as for the \citet{Singer2016} ansatz, giving a probability distribution for each line of sight. Our DPGMM is an alternative method for post-processing to produce three-dimensional localizations; below we show that it is effective, and a comparison of techniques for gravitational-wave source localization is left for future work. 

\section{Results}\label{sec:results}

In this section, we describe our findings for the localization of BNSs. We begin by verifying that our reconstructed posteriors are well calibrated (\secref{PP}). Then, we describe results for the (two-dimensional) sky-area analysis, before concluding with the full three-dimensional position results. A discussion of the implication of our results for multimessenger astronomy is given in \secref{conclusion}.

We report values for the credible regions and volumes, as well as the area or volume that would be searched (with a greedy algorithm) before discovering the true location \citep[cf.][]{Sidery2014}. The credible region $\mathrm{CR}_P$ is the smallest sky area that encompasses a total posterior probability $P$,
\begin{equation}
\mathrm{CR}_P = \min \left\{ A: \int_A \mathrm{d}\boldsymbol{\Omega}\, p(\boldsymbol{\Omega}) = P\right\},
\label{eq:CR}
\end{equation}
where $p(\boldsymbol{\Omega})$ is the posterior probability density over sky position $\boldsymbol{\Omega} = \{\alpha,\delta\}$, and $A$ is the sky area integrated over. The credible volume $\mathrm{CV}_P$ is the three-dimensional equivalent including distance too. We also use the distance credible interval $\mathrm{CI}_P$, which we define to be the central (equal-tailed) interval which contains probability $P$ \citep{Aasi2013}. The searched area $A_\ast$ is the size of the smallest credible region that includes the true location; the searched volume $V_\ast$ is the smallest credible volume that does the same. The sizes of credible regions and volumes indicate the precision of our parameter estimates, whereas the searched areas and volumes fold in the accuracy too.\footnote{For electromagnetic follow-up, the searched area would be the minimal area of the sky that a telescope would need to cover, starting from the most probable point, before imaging the true location. However, it may not be possible to immediately identify a transient as the counterpart to a gravitational-wave signal; therefore, a larger area may be covered in practice to avoid false identifications. Additionally, the need to tile with a finite field-of-view telescope can further increase the actual area searched.}

\subsection{Calibration}\label{sec:PP}

\begin{figure*}
	\centering
	\subfloat[][]{\includegraphics[width=0.65\columnwidth]{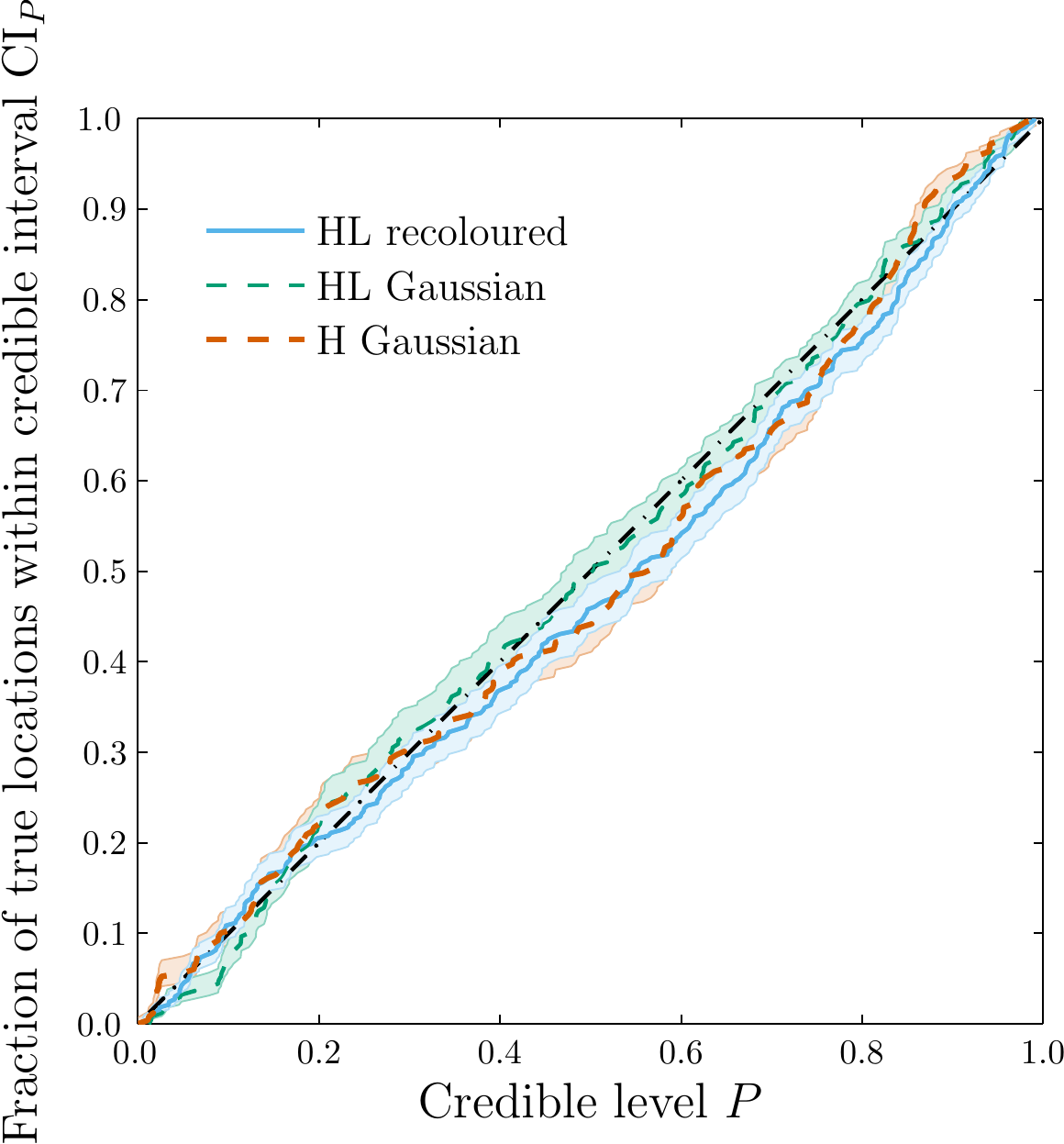}} \quad
	\subfloat[][]{\includegraphics[width=0.65\columnwidth]{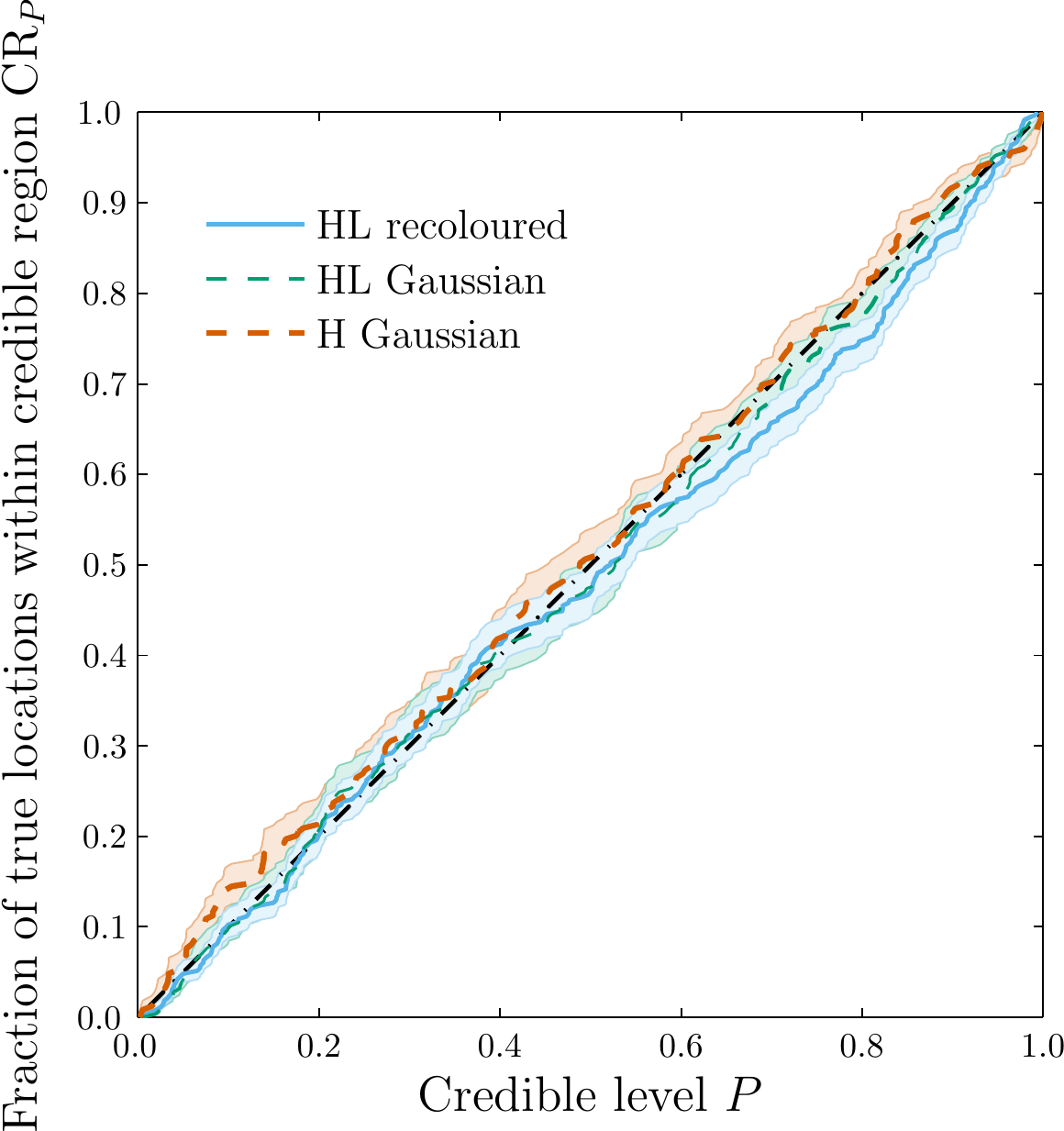}} \quad
	\subfloat[][]{\includegraphics[width=0.65\columnwidth]{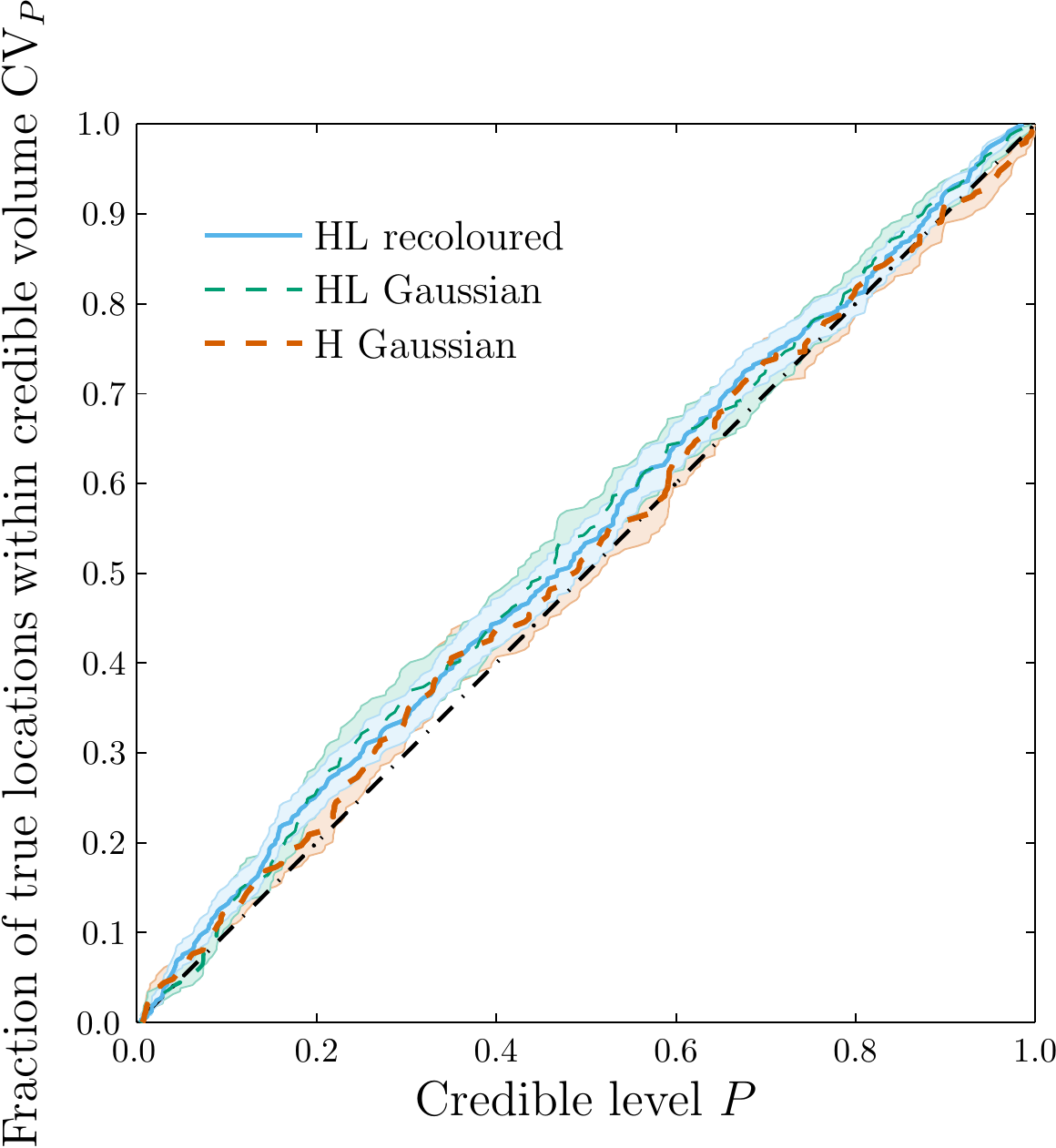}}
	\caption{Fraction of true locations found within a (a) distance credible interval, (b) sky-area credible region or (c) credible volume as a function of encompassed posterior probability. Results with simulated Gaussian noise are indicated by the dashed lines, results using recoloured S6 noise are indicated by the solid line, and the expected distribution is indicated by the dot--dashed diagonal line. The $68\%$ confidence interval for the cumulative distribution is enclosed by the shaded regions, this accounts for sampling errors and is estimated from a beta distribution \citep{Cameron2011}.} 
	\label{fig:pp}
\end{figure*}

To verify the self-consistency of results, we calculate the fraction of events that are located within the credible region or volume at a given probability. We expect that a proportion $P$ are found within $\mathrm{CI}_P$, $\mathrm{CR}_P$ or $\mathrm{CV}_P$ \citep{Cook2006}. A difference could arise if our prior does not match the injected distribution, but that should not be an issue here.\footnote{Our priors do agree with the injected distributions, and the posterior distributions have been previously verified for sky area and distance (but not volume) in \citet{Berry2014}.} \Figref{pp} shows the fraction of events found within a given $\mathrm{CI}_P$, $\mathrm{CR}_P$ and $\mathrm{CV}_P$ as a function of $P$; shown are results for three datasets, the HL Gaussian and HLV Gaussian results from \citet{Singer2014} and the HL recoloured results from \citet{Berry2014}.
Since the one-dimension distance and two-dimensional sky position probability distributions are constructed by marginalising the three-dimensional position probability distribution, the $\mathrm{CI}_P$, $\mathrm{CR}_P$ and $\mathrm{CV}_P$ results are not independent. Using a Kolmogorov--Smirnov (KS) test \citep[section 9.5]{DeGroot1975} to compare the expected and recovered distributions yields $p$-values of $0.09$, $0.72$ and $0.21$ for the HL recoloured, HL Gaussian and HLV Gaussian distances; $0.15$, $0.15$ and $0.62$ for the HL recoloured, HL Gaussian and HLV Gaussian sky areas, and $0.83$, $0.94$ and $0.58$ for the HL recoloured, HL Gaussian and HLV Gaussian volumes respectively. None of the distributions show any significant deviations away from the expected results. The posteriors appear to be well calibrated.

\subsection{Comparison with kernel density estimation}

As a further consistency check, we can compare sky area results generated using the DPGMM to those from KDE as used in \citet{Singer2014} and \citet{Berry2014}. This allows us to verify that both methods agree on an event-by-event basis. To summarize the variation in sky areas computed in different analyses, we use the log ratio \citep{Grover2013,Farr2015}
\begin{equation}
\mathcal{R}_A = \log_{10}\left(\frac{A\super{DP}}{A\super{KDE}}\right),
\end{equation}
where $A\super{DP}$ is a credible region or the searched area as determined by the DPGMM and $A\super{KDE}$ is the same quantity from the KDE. The log ratio is zero when both agree.

We find there is a scatter in the log ratio around zero, as summarised in \tabref{log-ratio}. The DPGMM results are more conservative on average, being $\sim10^{0.05} \simeq 1.1$ times larger than the KDE results. There is the largest difference in the HLV Gaussian results. This may be a consequence of these runs having a low number of (independent) posterior samples: the median number of posterior samples is $1000$ whereas the median number is $8600$ for both of the HL sets. Using a smaller set of posterior samples leads to less accurate estimates for the sky localization. The sky localization areas from the two approaches agree within the typical uncertainty of $\sim10\%$. 
\begin{table*}
	\centering
	\caption{Comparison of sky-localization areas produced using the DPGMM to those produced using KDE. The mean and standard deviation of the log ratio for the $50\%$ credible region $\mathrm{CR}_{0.5}$, the $90\%$ credible region $\mathrm{CR}_{0.9}$ and the searched area $A_\ast$ are listed for each method.}
	\label{tab:log-ratio}
	\begin{tabular}{l D{.}{.}{1.3} D{.}{.}{1.3} D{.}{.}{1.3} D{.}{.}{1.3} D{.}{.}{1.3} D{.}{.}{1.3} }
		\hline
		\multicolumn{1}{c}{Log} & \multicolumn{2}{c}{HL recoloured} & \multicolumn{2}{c}{HL Gaussian} & \multicolumn{2}{c}{HLV Gaussian} \\
		\multicolumn{1}{c}{ratio} & \multicolumn{1}{c}{Mean} & \multicolumn{1}{c}{Standard deviation} & \multicolumn{1}{c}{Mean} & \multicolumn{1}{c}{Standard deviation} & \multicolumn{1}{c}{Mean} & \multicolumn{1}{c}{Standard deviation} \\
		\hline
		$\mathcal{R}_{\mathrm{CR}_{0.5}}$ & 0.007 & 0.129 & 0.017 & 0.120 & 0.058 & 0.197 \\
		$\mathcal{R}_{\mathrm{CR}_{0.9}}$ & 0.047 & 0.135 & 0.045 & 0.134 & 0.072 & 0.192 \\
		$\mathcal{R}_{A_\ast}$            & 0.066 & 0.361 & 0.095 & 0.376 & 0.020 & 0.495 \\
		\hline
	\end{tabular}
\end{table*}

We do not expect perfect agreement between the approaches, since the DPGMM builds a three-dimensional probability distribution and projects this down to calculate sky areas whereas the KDE directly computes sky areas. We expect the KDE to perform better, since it especially designed to compute two-dimensional credible regions, and this is the case.

\subsection{Measurement uncertainty}

\subsubsection{Sky area}

\begin{figure*}
	\centering
	\subfloat[][]{\includegraphics[width=0.7525\columnwidth]{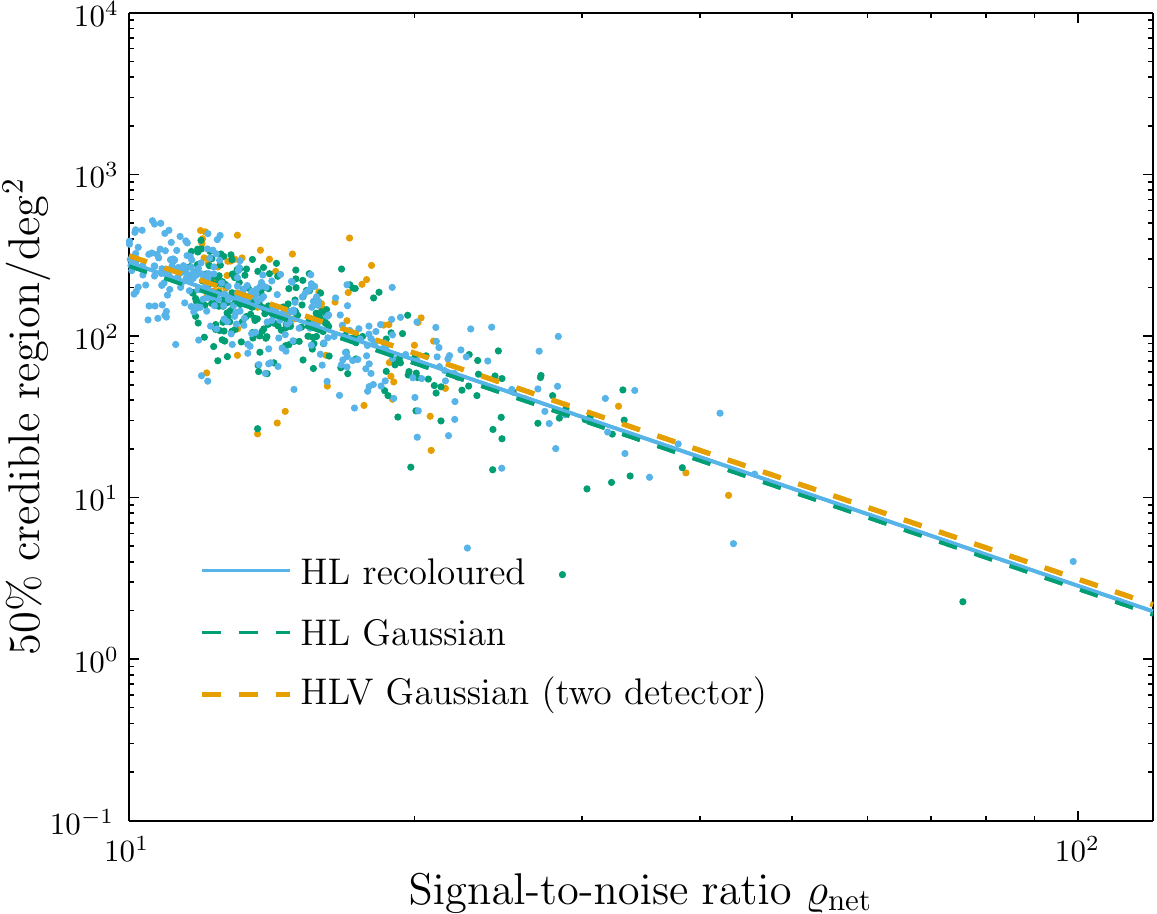}} \quad
	\subfloat[][]{\includegraphics[width=0.9325\columnwidth]{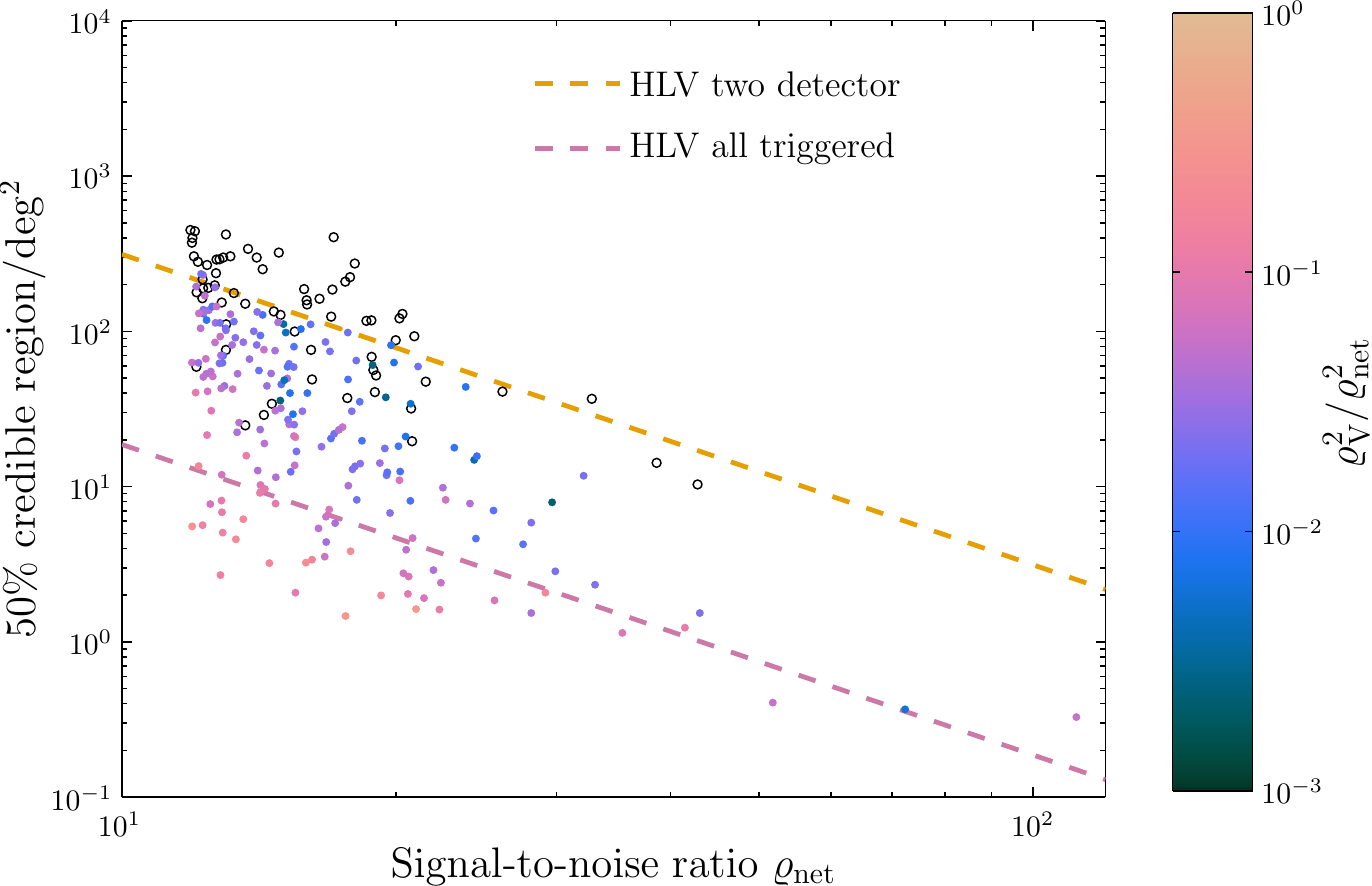}\label{fig:sky-50-3}} \\
	\subfloat[][]{\includegraphics[width=0.7525\columnwidth]{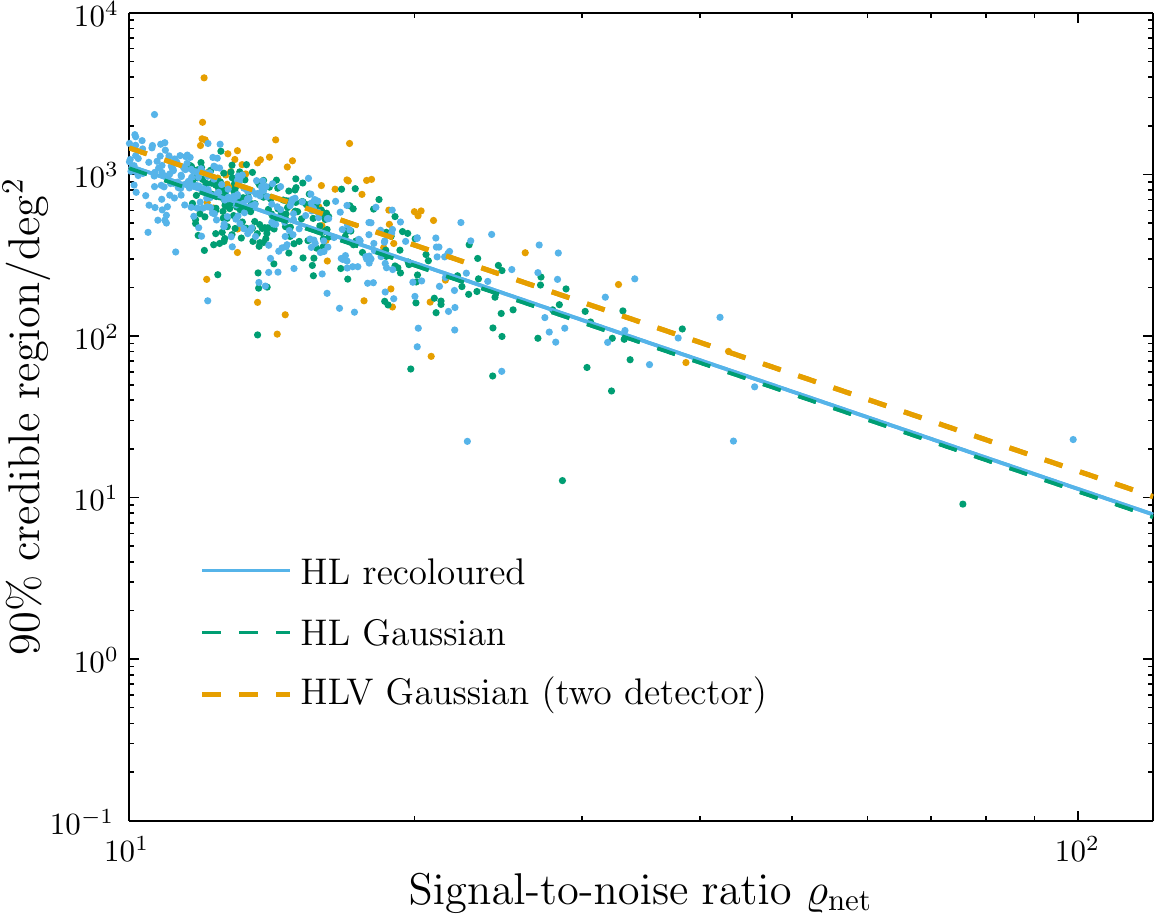}} \quad
	\subfloat[][]{\includegraphics[width=0.9325\columnwidth]{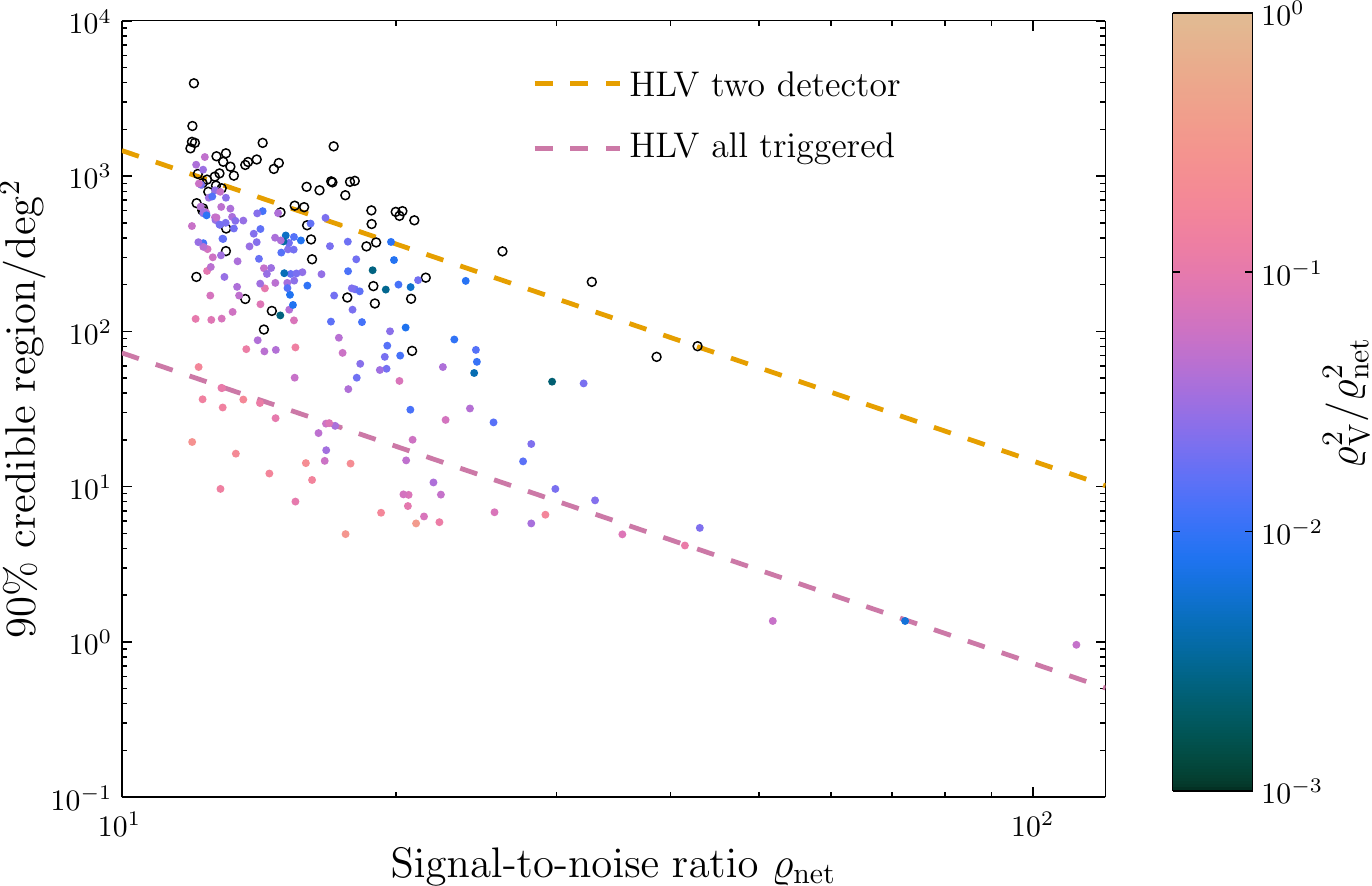}\label{fig:sky-90-3}} 
	\caption{Sky-localization areas as a function of SNR $\varrho\sub{net}$. The left column shows two-detector results and the right column shows all HLV scenario results; the top row shows the $50\%$ credible region $\mathrm{CR}_{0.5}$ and the bottom row shows the $90\%$ credible region $\mathrm{CR}_{0.9}$. Individual results are indicated by points and we include fiducial best-fit lines assuming that the area $A \propto \varrho\sub{net}^{-2}$. The left column shows both HL sets of results and the HLV results where only two detectors are operation, each has its own best-fit line. The HLV two-detector results are also shown in the right column, indicated by the open points, the three-detector results are colour-coded by the fraction of the SNR (squared) from AdV; the lines are fits to the two-detector network runs and those three-detector network runs loud enough to trigger in all detectors.} 
	\label{fig:sky-snr}
\end{figure*}

Having established that the DPGMM produces sensible results, we now present results for measurement accuracies. We begin by looking at sky-localization, as a final consistency check.
The sky-localization precision depends upon the SNR, scaling as $\varrho\sub{net}^{-2}$ \citep{Fairhurst2009,Berry2014}. We check this relationship in \figref{sky-snr}, where we plot credible regions versus SNR for the two-detector and three-detector networks. Unlike previous analyses in \citet{Berry2014} and \citet{Farr2015}, we do not use the SNR reported by the detection pipeline, but the SNR as determined by the maximum of the likelihood, $\mathcal{L} \sim \exp(-\varrho^2/2)$, found by \textsc{LALInference}. This is necessary as we consider events for HLV where there is no trigger (which requires a single-detector SNR of $4$), and hence no contribution to the \textsc{GstLAL}'s network SNR, from AdV, which is less sensitive than the aLIGO instruments. With a two-detector network, the scaling with SNR changes little between the HL and HLV scenarios (or when considering different combinations of two detectors for HLV); there is slightly worse performance for HLV as a result of a decrease in frequency bandwidth at a given SNR \citep{Singer2014}. In the HLV scenario, the big change comes from the introduction of a third detector. The improvement from the third detector is continuous \citep{Abbott2017e}, ranging from providing negligible additional information to a reduction in sky area (at a given network SNR) by a factor of $\sim16$; this is heuristically illustrated by the fraction of the SNR from AdV $\varrho\sub{V}/\varrho\sub{net}$, indicated by the colour-coding in \figref{sky-50-3} and \figref{sky-90-3}.

\subsubsection{Volume}

\begin{figure}
	\centering
	\subfloat[][]{\includegraphics[width=0.8\columnwidth]{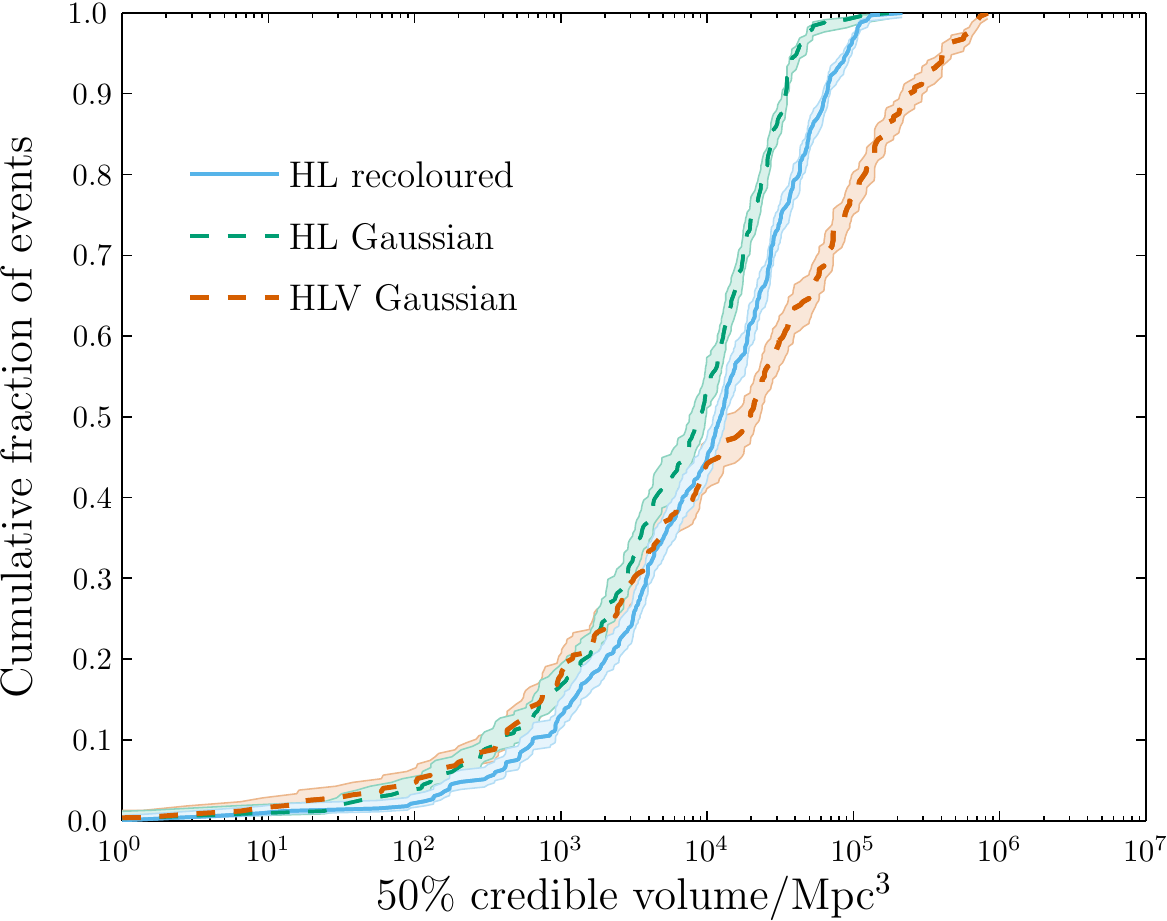}} \\
	\subfloat[][]{\includegraphics[width=0.8\columnwidth]{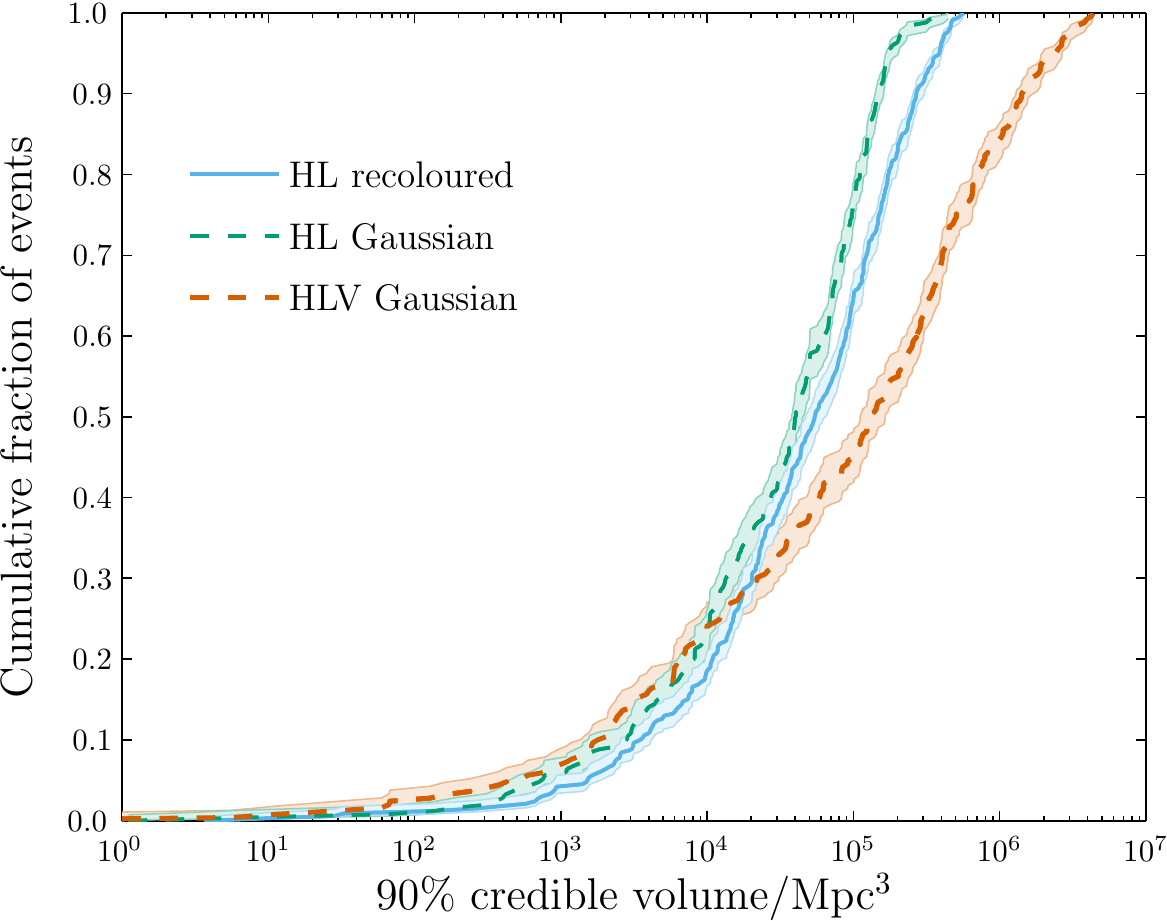}} \\
	\subfloat[][]{\includegraphics[width=0.8\columnwidth]{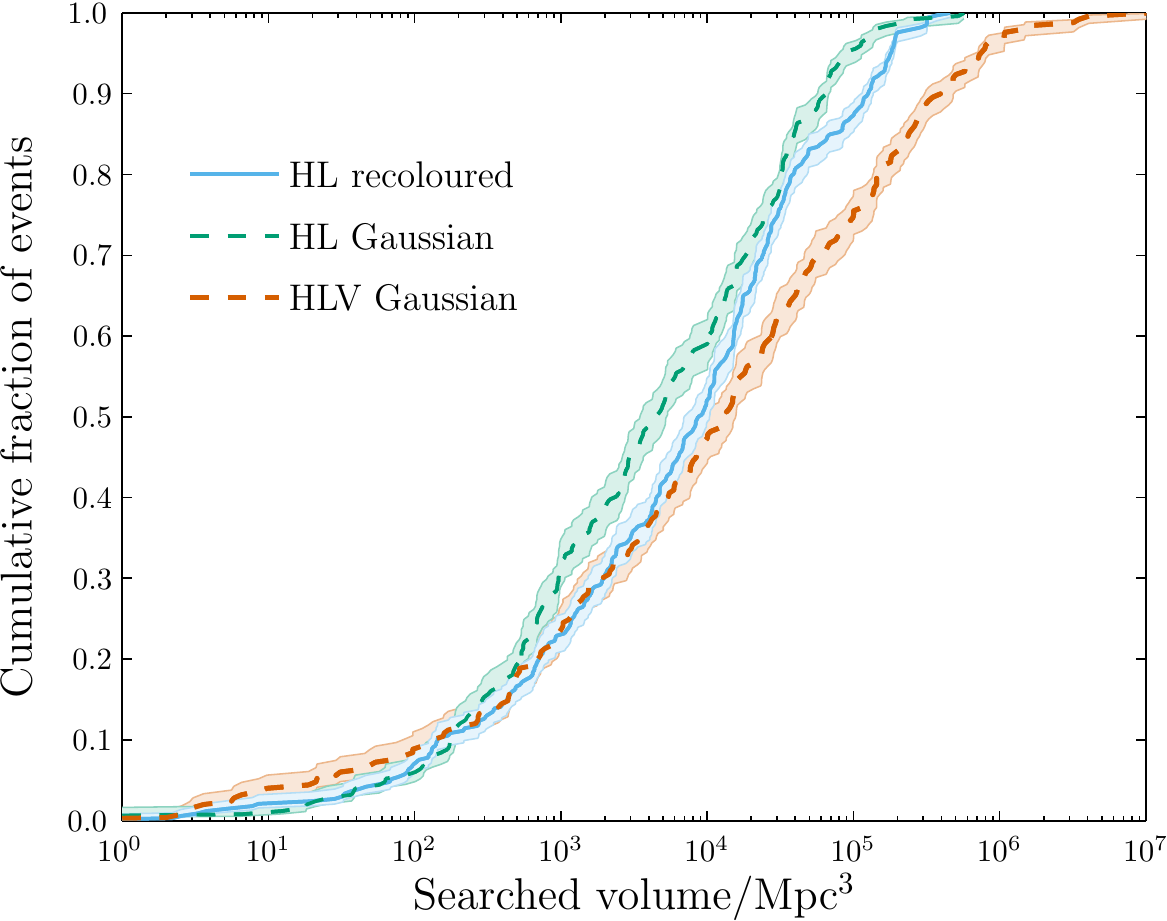}}
	\caption{Cumulative fractions of events with localization volumes smaller than the abscissa value. The top panel shows the $50\%$ credible volume $\mathrm{CV}_{0.5}$, the middle shows the $90\%$ credible volume $\mathrm{CV}_{0.9}$ and the bottom shows the searched volume $V_\ast$. The $68\%$ confidence interval for the cumulative distribution is enclosed by the shaded regions; this does not include the inherent uncertainty in the volume estimates.} 
	\label{fig:volume-cumulative}
\end{figure}

Finally, we consider the full three-dimensional localization. The cumulative distributions of localization volumes, as constructed from our DPGMM, are shown in \figref{volume-cumulative}. Statistics summarising these distributions are given in \tabref{volume-fraction} and \tabref{volume-median}. The three sets of results are similar; the volumes for the HL recoloured results are slightly larger than the HL Gaussian results on account of the additional low SNR events, and the HLV Gaussian results are \emph{larger} still as the increased detector sensitivity allows us to detect sources at a greater distance.\footnote{The median true distances of detections are $50.1~\mathrm{Mpc}$, $47.8~\mathrm{Mpc}$ and $97.0~\mathrm{Mpc}$ for the HL recoloured, HL Gaussian and HLV Gaussian sets respectively.}
\begin{table*}
	\centering
	\caption{Fractions of events with localization volumes smaller than a given size, and the fraction of searched volumes which contain fewer than the given number of galaxies in the GLADE catalogue \citep{Dalya2018}. A dash (---) is used for fractions less than $0.01$.}
	\label{tab:volume-fraction}
	\begin{tabular}{l l D{.}{.}{1.2} D{.}{.}{1.2} D{.}{.}{1.2} D{.}{.}{1.2} D{.}{.}{1.2}}
		\hline
		\multicolumn{2}{c}{Volume or} & \multicolumn{1}{c}{HL}         & \multicolumn{1}{c}{HL}       & \multicolumn{1}{c}{Two-detector} & \multicolumn{1}{c}{Three-detector} & \multicolumn{1}{c}{All HLV}   \\
                \multicolumn{2}{c}{no.\ of galaxies} & \multicolumn{1}{c}{recoloured} & \multicolumn{1}{c}{Gaussian} & \multicolumn{1}{c}{HLV Gaussian}        & \multicolumn{1}{c}{HLV Gaussian} & \multicolumn{1}{c}{Gaussian}  \\
		\hline
		\multirow{6}{*}{$\displaystyle\frac{\mathrm{CV}_{0.5}}{\mathrm{Mpc^3}} \leq$} 
			& $10$   & 0.01                    & \multicolumn{1}{c}{---} & \multicolumn{1}{c}{---} & 0.02                    & 0.02 \\
			& $10^2$ & 0.02                    & 0.04                    & 0.01                    & 0.06                    & 0.05 \\
			& $10^3$ & 0.13                    & 0.17                    & 0.06                    & 0.22                    & 0.18 \\
			& $10^4$ & 0.45                    & 0.54                    & 0.31                    & 0.49                    & 0.44 \\
			& $10^5$ & 0.97                    & 0.99                    & 0.66                    & 0.82                    & 0.78 \\
			& $10^6$ & 1.00                    & 1.00                    & 1.00                    & 1.00                    & 1.00 \\
		\hline
		\multirow{6}{*}{$\displaystyle\frac{\mathrm{CV}_{0.9}}{\mathrm{Mpc^3}} \leq$} 
			& $10$   & \multicolumn{1}{c}{---} & \multicolumn{1}{c}{---} & \multicolumn{1}{c}{---} & \multicolumn{1}{c}{---} & \multicolumn{1}{c}{---} \\
			& $10^2$ & 0.01                    & 0.01                    & \multicolumn{1}{c}{---} & 0.04                    & 0.03 \\
			& $10^3$ & 0.04                    & 0.06                    & 0.01                    & 0.09                    & 0.07 \\
			& $10^4$ & 0.19                    & 0.23                    & 0.14                    & 0.27                    & 0.24 \\
			& $10^5$ & 0.65                    & 0.77                    & 0.31                    & 0.50                    & 0.45 \\
			& $10^6$ & 1.00                    & 1.00                    & 0.78                    & 0.87                    & 0.84 \\
		\hline
		\multirow{6}{*}{$\displaystyle\frac{V_\ast}{\mathrm{Mpc^3}} \leq$}            
			& $10$   & 0.02                    & 0.01                    & \multicolumn{1}{c}{---} & 0.05                    & 0.04 \\
			& $10^2$ & 0.07                    & 0.06                    & 0.05                    & 0.10                    & 0.09 \\
			& $10^3$ & 0.23                    & 0.32                    & 0.13                    & 0.27                    & 0.24 \\
			& $10^4$ & 0.52                    & 0.59                    & 0.36                    & 0.52                    & 0.47 \\
			& $10^5$ & 0.87                    & 0.95                    & 0.70                    & 0.77                    & 0.75 \\
			& $10^6$ & 1.00                    & 1.00                    & 0.98                    & 0.97                    & 0.97 \\
		\hline
		\multirow{5}{*}{$\displaystyle{n\super{G}_\ast} \leq$}            
			& $1$    & 0.03                    & 0.03                    & 0.02                    & 0.08                    & 0.07 \\
			& $10$   & 0.20                    & 0.24                    & 0.15                    & 0.26                    & 0.23 \\
			& $10^2$ & 0.51                    & 0.57                    & 0.30                    & 0.52                    & 0.46 \\
			& $10^3$ & 0.88                    & 0.95                    & 0.73                    & 0.80                    & 0.78 \\
			& $10^4$ & 1.00                    & 1.00                    & 1.00                    & 0.98                    & 0.98 \\
	
		\hline
	\end{tabular}
\end{table*}
\begin{table*}
	\centering
	\caption{Median localization volumes constructed using the DPGMM, and the median number of galaxies in the GLADE catalogue \citep{Dalya2018} within these volumes.}
	\label{tab:volume-median}
	\begin{tabular}{l  D{;}{\times}{3.3} D{;}{\times}{3.3} D{;}{\times}{3.3} D{;}{\times}{3.3} D{;}{\times}{3.3}}
		\hline
		                           & \multicolumn{1}{c}{HL}         & \multicolumn{1}{c}{HL}       & \multicolumn{1}{c}{Two-detector} & \multicolumn{1}{c}{Three-detector} & \multicolumn{1}{c}{All HLV}   \\
                \multicolumn{1}{c}{Median} & \multicolumn{1}{c}{recoloured} & \multicolumn{1}{c}{Gaussian} & \multicolumn{1}{c}{HLV Gaussian}  & \multicolumn{1}{c}{HLV Gaussian} & \multicolumn{1}{c}{Gaussian} \\
		\hline
		${\mathrm{CV}_{0.5}}/\mathrm{Mpc^3}$ & 1.2;10^4 & 8.9;10^3 & 5.2;10^4 & 1.3;10^4 & 2.0;10^4 \\
		${\mathrm{CV}_{0.9}}/\mathrm{Mpc^3}$ & 5.4;10^4 & 4.0;10^4 & 2.9;10^5 & 1.0;10^5 & 1.3;10^5 \\
		${V_\ast}/\mathrm{Mpc^3}$            & 8.7;10^3 & 4.4;10^3 & 2.9;10^4 & 9.1;10^3 & 1.3;10^4 \\
		\hline
		${n\super{G}_{0.5}}$               & 1.3;10^2 & 8.1;10^1 & 3.5;10^2 & 9.2;10^1 & 1.5;10^2 \\
		${n\super{G}_{0.9}}$               & 5.9;10^2 & 4.4;10^2 & 2.2;10^3 & 7.6;10^2 & 1.1;10^3 \\
		${n\super{G}_\ast}$                & 9.5;10^1 & 5.6;10^1 & 2.7;10^2 & 8.6;10^1 & 1.2;10^2 \\
		\hline
	\end{tabular}
\end{table*}

\begin{figure*}
	\centering
	\subfloat[][]{\includegraphics[width=0.7525\columnwidth]{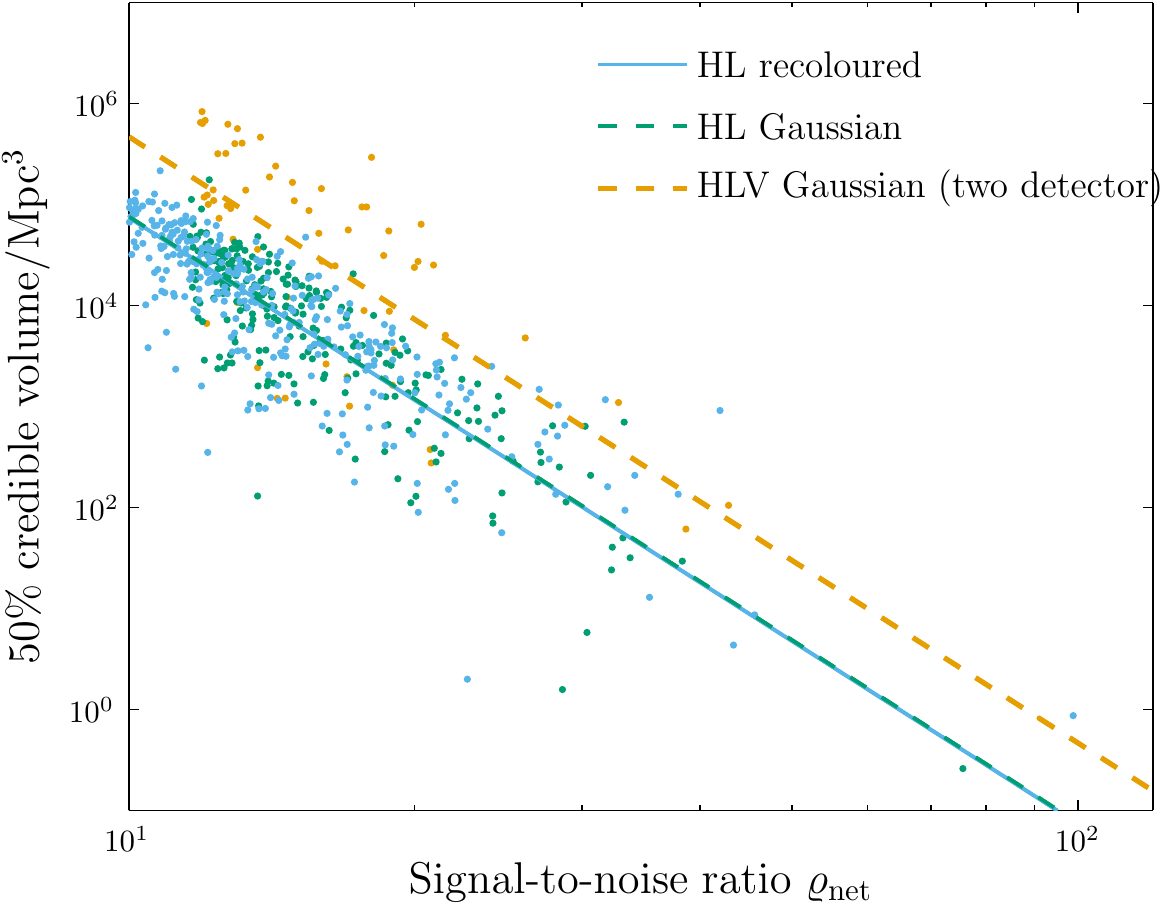}} \quad
	\subfloat[][]{\includegraphics[width=0.9325\columnwidth]{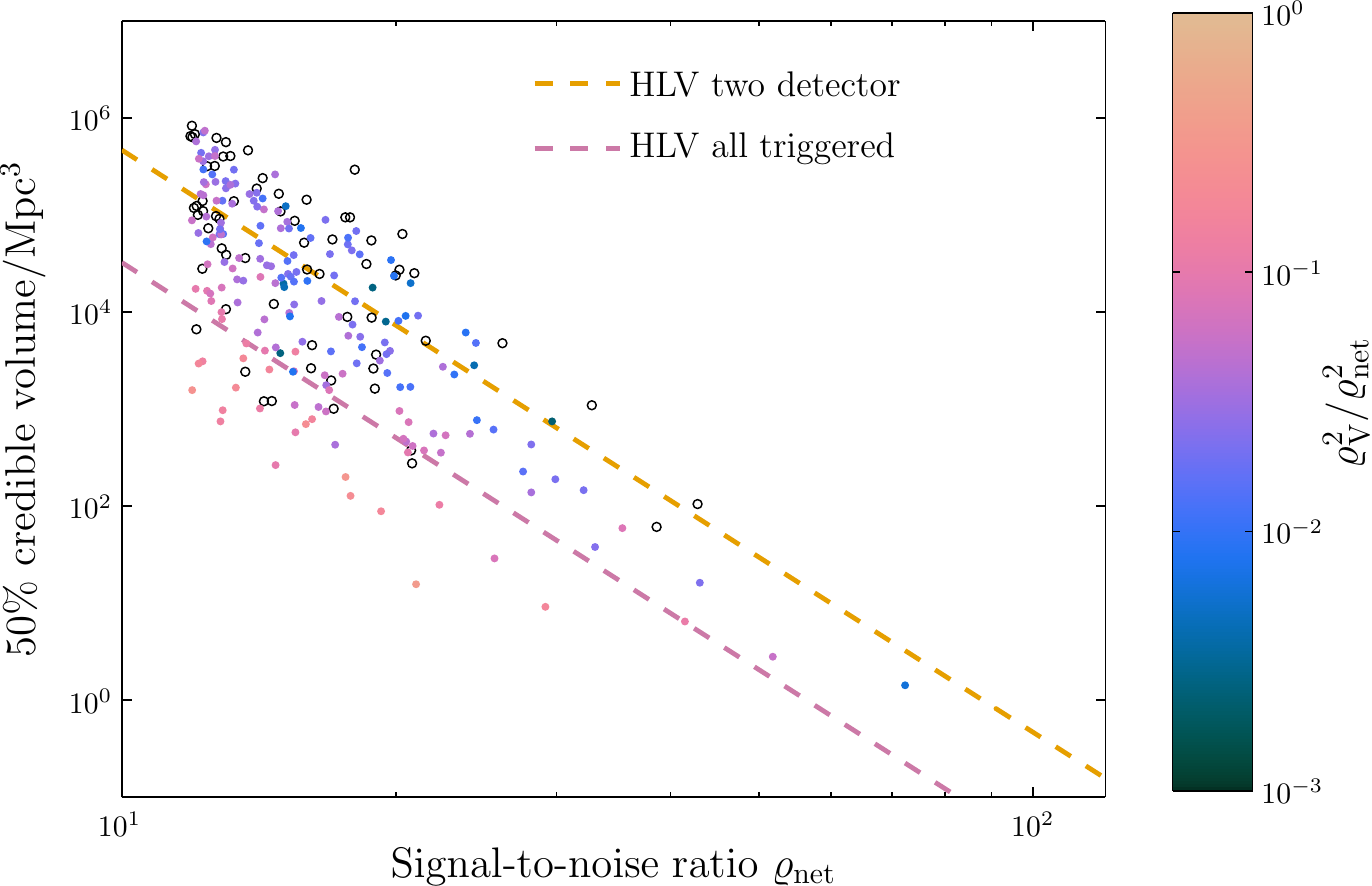}\label{fig:volume-50-3}} \\
	\subfloat[][]{\includegraphics[width=0.7525\columnwidth]{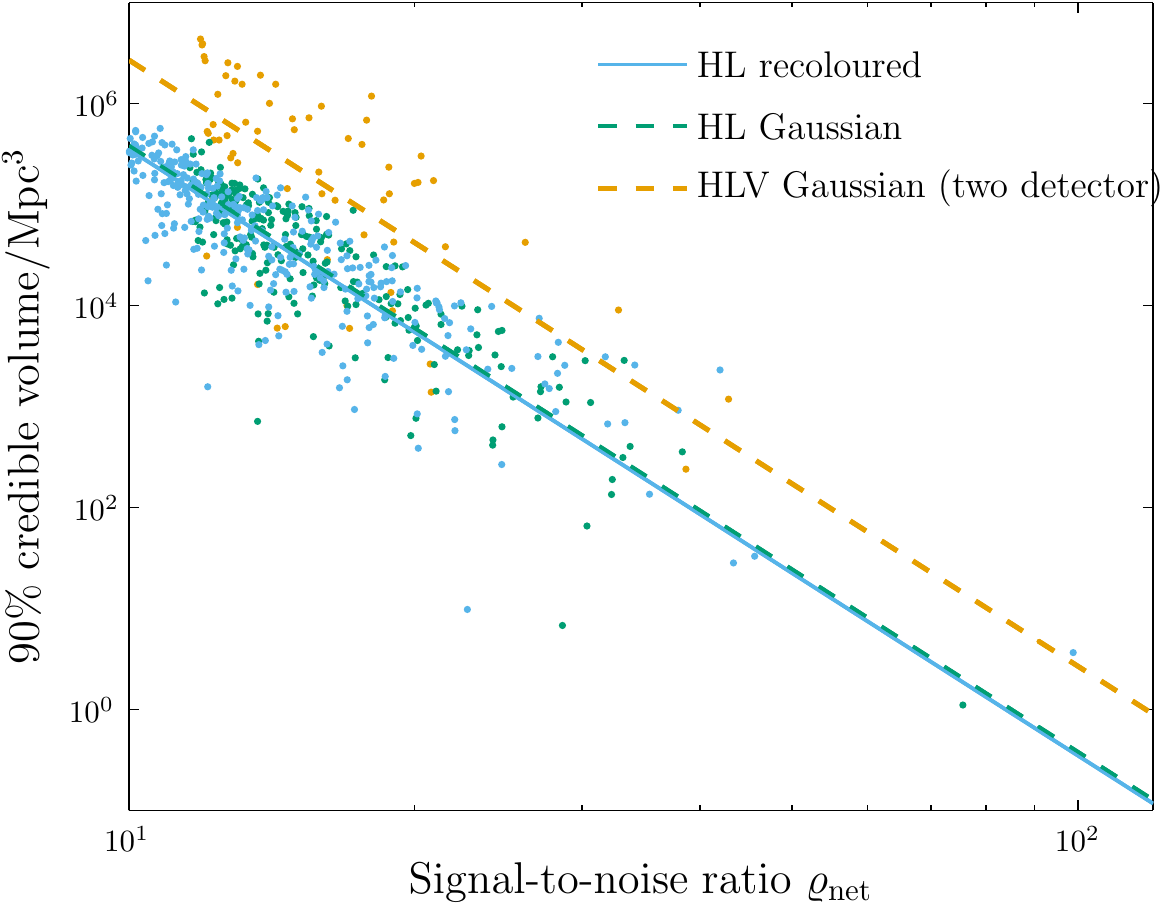}} \quad
	\subfloat[][]{\includegraphics[width=0.9325\columnwidth]{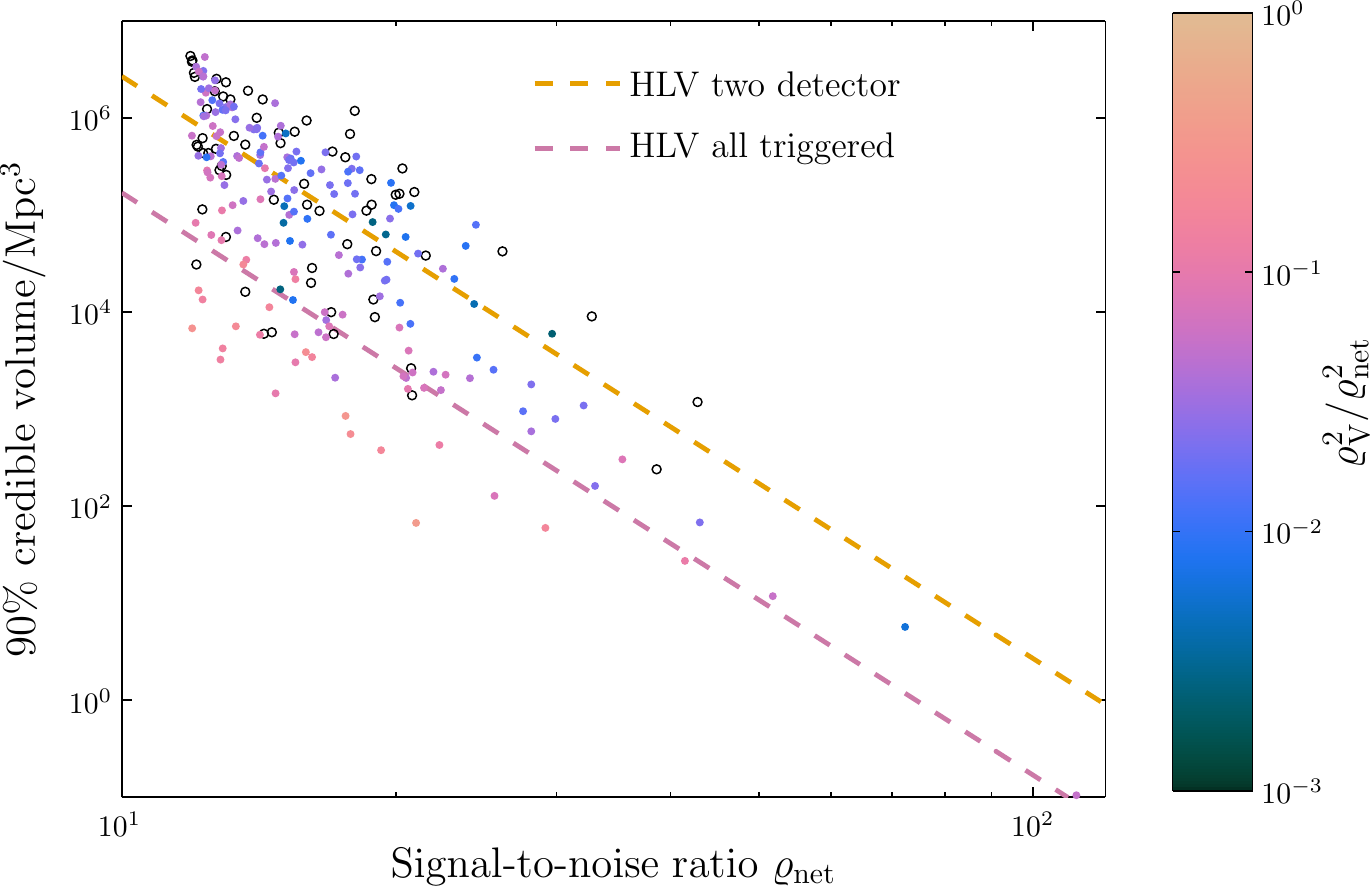}\label{fig:volume-90-3}} 
	\caption{Localization volumes as a function of SNR $\varrho\sub{net}$. The left column shows two-detector results and the right column shows all HLV scenario results; the top row shows the $50\%$ credible volume $\mathrm{CV}_{0.5}$ and the bottom row shows the $90\%$ credible volume $\mathrm{CV}_{0.9}$. Individual results are indicated by points and we include fiducial best-fit lines assuming that the volume $V \propto \varrho\sub{net}^{-6}$. The left column shows both HL sets of results and the HLV results where only two detectors are operation, each has its own best-fit line. The HLV two-detector results are also shown in the right column, indicated by the open points, the three-detector results are colour-coded by the fraction of the SNR (squared) from AdV; the lines are fits to the two-detector network runs and those three-detector network runs loud enough to trigger in all detectors.} 
	\label{fig:volume-snr}
\end{figure*}

The three-dimensional localization also depends upon the SNR. The uncertainty in the three-dimensional location can be estimated as
\begin{equation}
\Delta V \sim D^2 \Delta D \Delta A,
\end{equation}
where $\Delta D$ and $\Delta A$ are the uncertainty on the distance and sky location respectively. The distance is inversely proportional to the signal amplitude (keeping all other parameters fixed) and hence $D \propto \varrho\sub{net}^{-1}$; from a Fisher-matrix analysis, we expect that the fractional error in the distance is inversely proportional to the SNR $\Delta D/D \propto \varrho\sub{net}^{-1}$ \citep{Cutler1994}, and we have seen that $\Delta A \propto \varrho\sub{net}^{-2}$ (\figref{sky-snr}). Combining these, we expect that
\begin{equation}
\Delta V \propto \frac{1}{\varrho\sub{net}^6}.
\end{equation}
The credible volumes versus SNR are plotted in \figref{volume-snr} for the two-detector and three-detector networks. The trends are roughly as expected; there is significant scatter because the SNR also depends upon other source properties such as the binary inclination and the sky position relative to the detectors. We see that, although on average the HLV scenario localization is worse than in the HL scenario, when we only consider events with significant SNR in all three detectors, the localization is better than in HL \citep[cf.][]{Veitch2012}. Adding a third detector in the HLV scenario can improve localization by (on average) a factor of $\sim15$.

\subsection{Applications for electromagnetic follow-up}\label{sec:EM}

Gravitational-wave sky localizations can be large \citep[e.g.,][]{Abbott2016a}, making the prompt search for an electromagnetic counterpart difficult. The extra information inherent in a three-dimensional localization can help optimise this search. For example, astronomers could choose to prioritise areas of the sky where the source is more probable to be close by and hence appear brighter, or adjust exposure times such that times are longer where the distance is probably larger and shorter where the distance is probably smaller. A significant improvement is potentially possible by looking for counterparts that are coincident with galaxies, as opposed to searching blindly \citep[e.g.,][]{Nissanke2012,Hanna2014,Blackburn2015,Gehrels2015,Singer2016}, and this strategy was followed by several teams searching for counterparts to GW170817 using the three-dimensional localization provided by the LVC \citep{Abbott2017b}.

Using our DPGMM, it is simple to correlate our three-dimensional posterior probability distributions with galaxy catalogues to produce a list of most probable galaxies. This only takes a few minutes to calculate; 
since we do not have to evaluate the DPGMM on a grid, it is \emph{quicker} than producing credible volumes. We make use of the Galaxy List for the Advanced Detector Era (GLADE) catalogue \citep{Dalya2015,Dalya2018}.\footnote{Available from \href{http://aquarius.elte.hu/glade/}{aquarius.elte.hu/glade/}.} This is constructed from the Gravitational Wave Galaxy Catalogue \citep[GWCC;][]{White2011}, the Two~Micron All-Sky Survey Extended Source Catalog \citep[2MASS XSC;][]{Skrutskie2006}, the Two~Micron All-Sky Survey Photometric Redshift catalog \citep[2MPZ;][]{Bilicki2014}, and HyperLeda catalogue \citep{Makarov2014}; it contains $\sim2,000,000$ galaxies, and is estimated to be complete to $73~\mathrm{Mpc}$ and $53\%$ complete at $300~\mathrm{Mpc}$. 

\begin{figure*}
	\centering
	\subfloat[][]{\includegraphics[width=1.865\columnwidth]{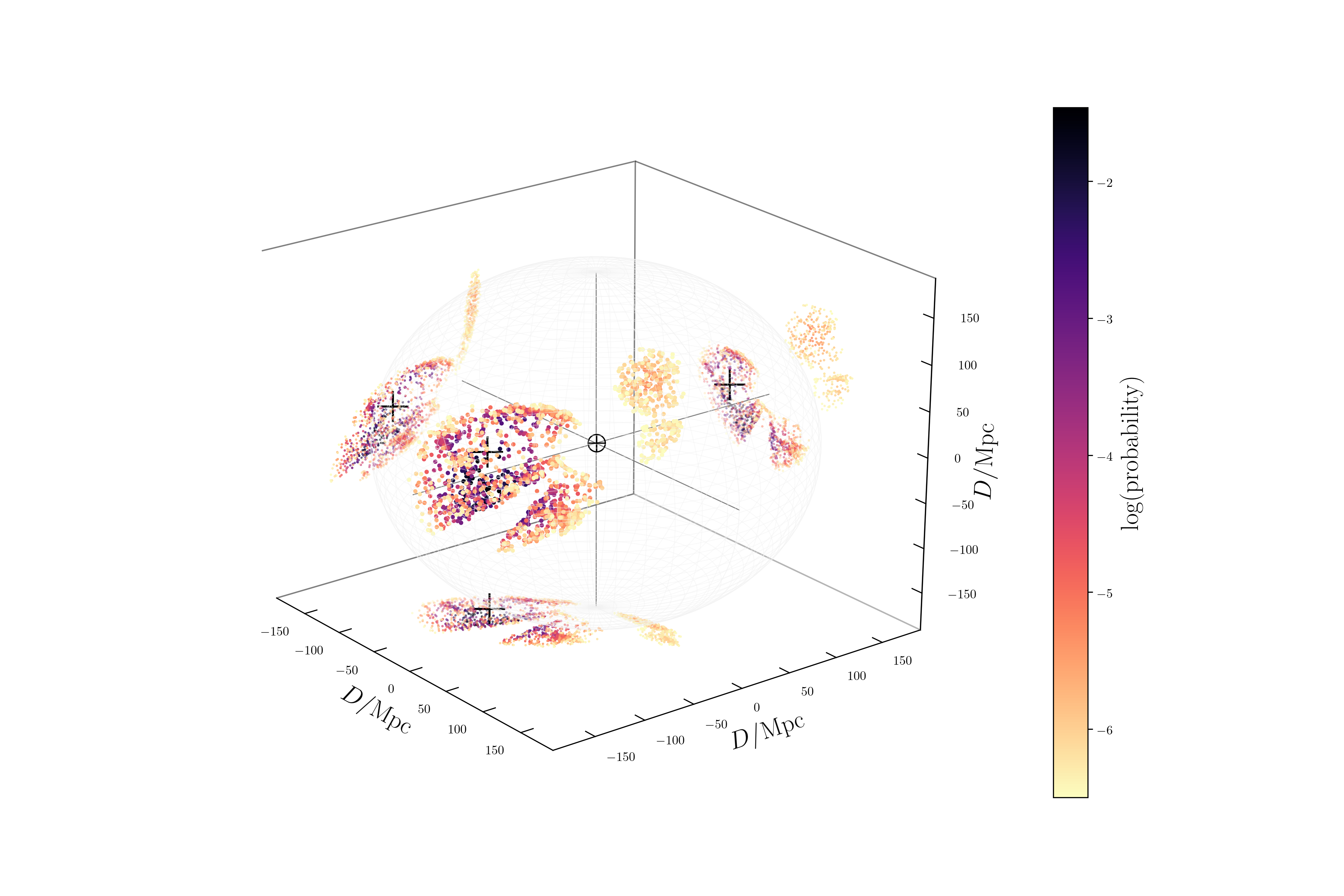}}\\
	\subfloat[][]{\includegraphics[width=1.865\columnwidth]{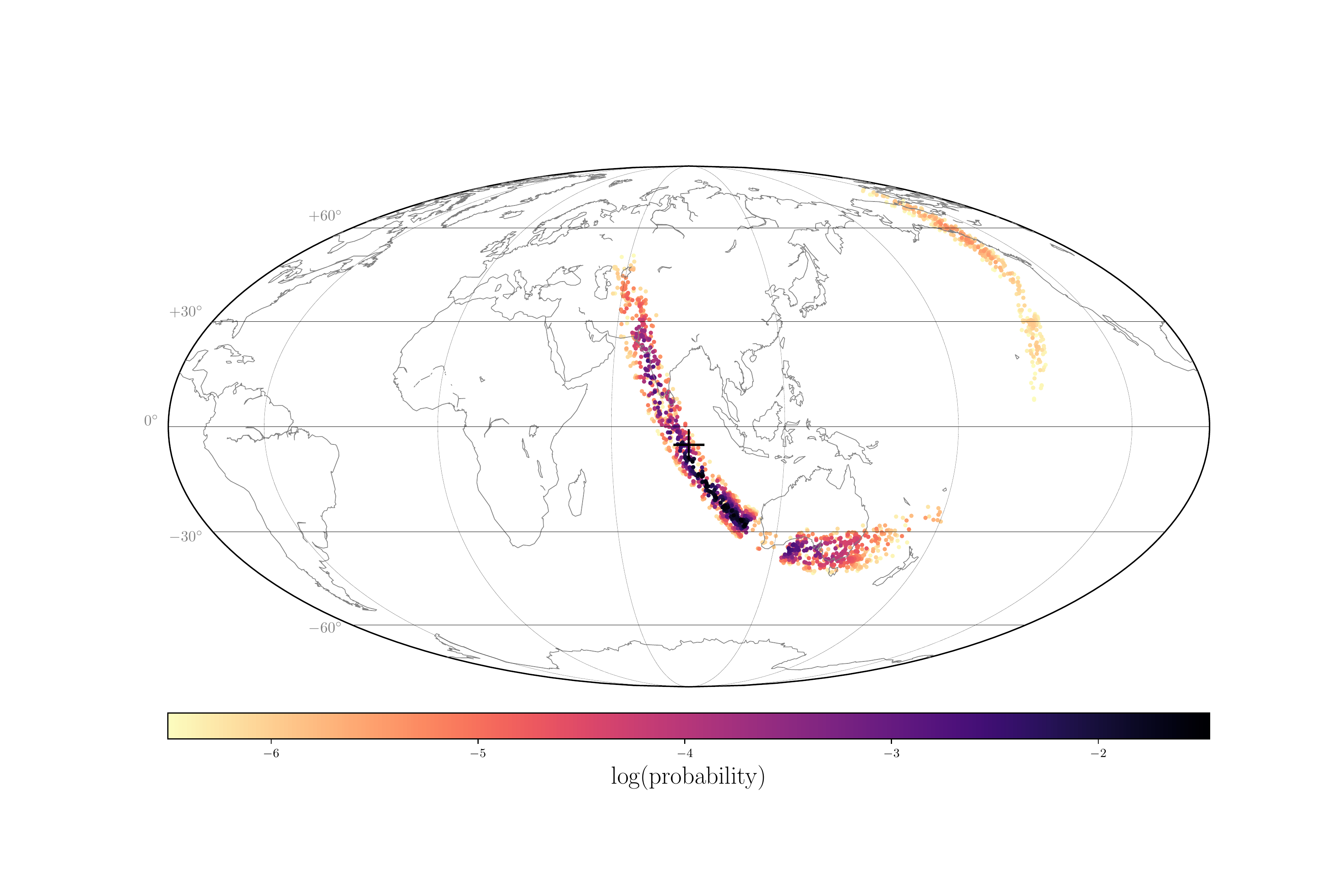}}
	\caption{Example posterior distribution for possible galaxy hosts from the GLADE catalogue \citep{Dalya2018}. We show the full three-dimensional scatter plot (top) and its projection onto the plane of the sky (bottom). In both panels, galaxies are colour-coded according to the (log) posterior probability of being the host of source, and we show galaxies with the $90\%$ credible volume. In the three-dimensional plot, we show the projections along the axes directions to aid in recognising the three dimensional shape of the posteriors. The black crosses indicate the true source location. The gap in the larger branch of the distribution is due to the incompleteness of the catalogue in the direction of the plane of the Milky Way.} 
	\label{fig:galaxies}
\end{figure*}

As an example of the end data product of our analysis, \figref{galaxies} shows a DPGMM localization correlated with galaxies from the GLADE catalogue \citep{Dalya2018}. The full three-dimensional posterior distribution is shown in the top panel, and its projection onto the plane of the sky is shown in the bottom panel. These show the characteristic shapes of localizations; they are not simple blobs, but can form disjoint regions \citep[described as jacaranda seeds in][]{Singer2016}. From the two panels, we can see the benefit of the additional information gained by considering the three-dimensional localization, instead of only a two-dimensional localization; the probable distance range is not the same for all lines of sight.

The most probable galaxies provide a starting point for a counterpart search. Further refinements could be made, such as factoring in the stellar mass of the galaxies \citep[cf.][]{Nuttall2010}, potentially by using luminosity as a mass proxy \citep[e.g.,][]{Hanna2014,Fan2014,Arcavi2017a}.

In \tabref{volume-median}, we include the number of galaxies included in the GLADE catalogue within the credible volumes $\mathrm{CV}_{0.5}$ and $\mathrm{CV}_{0.9}$, and the searched volume $V_\ast$: $n\super{G}_{0.5}$, $n\super{G}_{0.9}$ and $n\super{G}_{\ast}$, respectively.\footnote{Since the original set of simulated signals were drawn uniformly in volume, rather than from a galaxy catalogue, we cannot identify a true host galaxy which must be imaged to find the source.} These are lower limits on the true number of galaxies, but provide estimates for the number of galaxies that would be searched using the catalogue, and following a greedy algorithm weighting the galaxies by probability from the three-dimensional localization. In \tabref{volume-fraction}, we give numbers quantifying the distribution of $n\super{G}_{\ast}$. The number of catalogue galaxies in the localization volumes are approximately consistent with a density of one galaxy per $100~\mathrm{Mpc^3}$. 

\section{Conclusions}\label{sec:conclusion}

We have explained how DPGMMs can be used for post-processing of parameter-estimation studies. This technique will be useful for a variety of inference problems within astrophysics. We have applied our approach to an example from gravitational-wave astronomy, reconstructing the three-dimensional location of a BNS using results from \textsc{LALInference}.

The era of gravitational-wave astronomy is here, and we need to understand how to extract the maximum amount of information from signals. Localization of BNS sources is important for multimessenger astronomy as it allows for cross-referencing with galaxy catalogues. This is beneficial when searching for an electromagnetic counterpart \citep{Nissanke2012,Hanna2014,Gehrels2015,Singer2016}, as for GW170817 \citep{Abbott2017b}, but is still useful when none is found, for example for measurements of the Hubble constant \citep{Schutz1986,DelPozzo2012,Chen2017}. The DPGMM three-dimensional localizations can be be used to find the most probable source galaxies within a matter of minutes of the \textsc{LALInference} analysis finishing, making it useful for prompt multimessenger follow-up activities.

We constructed localization volumes for a catalogue of BNS signals appropriate for the early operation of the advanced-detector era \citep{Singer2014,Berry2014,Farr2015}. We have verified that the three-dimensional localizations are well calibrated \citep[cf.][]{Cook2006,Sidery2014} and have confirmed that when distance is marginalised out, these volumes reduce to sky areas that are consistent with two-dimensional KDE results. Our credible volumes have the expected proportionality with SNR, scaling roughly $\propto \varrho\sub{net}^{-6}$.

Our results show that localizations for detections during early observing runs would be $\sim10^4$--$10^5~\mathrm{Mpc^3}$, corresponding to $\sim10^2$--$10^3$ potential host galaxies within the GLADE catalogue \citep{Dalya2018}. Approximately half of events have searched volumes which contain $10^2$ galaxies or fewer, and a few percent of events have searched volumes which contain a single galaxy. Since our results do not include the effects of calibration uncertainty, they would be lower bounds for any actual detections: for the (O1-like) HL recoloured data set, we find that the median $90\%$ credible volume is $5\times10^4~\mathrm{Mpc^3}$ and for the HL Gaussian data set it is $4\times10^4~\mathrm{Mpc^3}$; moving ahead to the (O2-like) HLV scenario, the median $90\%$ credible volume is $1\times10^5~\mathrm{Mpc^3}$ for the Gaussian data set. Greater sensitivity of the detectors means that we can detect signals from a greater distance and hence are sensitive to sources in a larger volume. However, localization does improve as further detectors are added to the network: the median $90\%$ credible volume in the HLV scenario for a two-detector network is $3\times10^5~\mathrm{Mpc^3}$ but for a three-detector network it is $1\times10^5~\mathrm{Mpc^3}$. The localization improves rapidly as the SNR of the signal increases, and the best localization occurs when there is significant SNR from each of the three detectors. Addition of further detectors, such as KAGRA \citep{Aso2013} or the proposed LIGO-India detector \citep{Unnikrishnan2013,Abbott2017e}, could further improve localization and the prospects of identifying a counterpart.

\section*{Acknowledgements}

The authors are grateful for useful suggestions from the CBC group of the LIGO Scientific and Virgo Collaborations; 
WDP thanks Neil Cornish, Tjonnie Li, Trevor Sidery and John Veitch for early suggestions and discussions, we thank Will Farr for discussions on localization algorithms, and thank Ilya Mandel, Jonathan Gair, Hannah Middleton, Ewan Cameron and the anonymous referee for comments on the manuscript. 
We thank the other authors of \citet{Singer2014} and \citet{Berry2014} for sharing the data for this work. We also would like to thank contributors of GLADE for making the catalogue publicly available, and especially Gergely D{\'a}lya for help with its documentation and use. This work was supported in part a by Leverhulme Trust research project grant and in part by the Science and Technology Facilities Council. 
LIGO was constructed by the California Institute of Technology and Massachusetts Institute of Technology with funding from the National Science Foundation and operates under cooperative agreement PHY-0757058. 
This work used computing resources of the LIGO Data Grid including: the Atlas computing cluster at the Albert Einstein Institute, Hannover; the LIGO computing clusters at Caltech, and the facilities of the Advanced Research Computing @ Cardiff (ARCCA) Cluster at Cardiff University.
We are grateful for computational resources provided by the Leonard E Parker Center for Gravitation, Cosmology and Astrophysics at University of Wisconsin-Milwaukee. 
Some results were produced using the post-processing tools of the \texttt{plotutils} library at \href{http://github.com/farr/plotutils}{github.com/farr/plotutils} and \texttt{skyarea} library at \href{http://github.com/farr/skyarea}{github.com/farr/skyarea}. The Dirichlet Process Gaussian-mixture Model is included as a module available from \href{http://github.com/thaines/helit/}{github.com/thaines/helit/} and our implementation for three-dimensional localization is available from \href{http://github.com/wdpozzo/3d\_volume}{github.com/wdpozzo/3d\_volume}. 
We thank GW150914, GW170104 and GW170817 for delaying the completion of this work.

\bibliographystyle{mnras}
\bibliography{DirichletPaper}

\bsp
\label{lastpage}
\end{document}